\documentclass[a4paper,fleqn,usenatbib]{mnras}

\usepackage{mathptmx}

\usepackage[T1]{fontenc}
\usepackage{ae,aecompl}

\usepackage{graphicx}	
\usepackage{float}
\graphicspath{{"figure/"}}  
\usepackage{amsmath}	
\usepackage{amssymb}	
\usepackage{cases}
\usepackage{tabularx}
\usepackage{booktabs}

\newcolumntype{L}[1]{>{\hsize=#1\hsize\raggedright\arraybackslash}X}%
\newcolumntype{R}[1]{>{\hsize=#1\hsize\raggedleft\arraybackslash}X}%
\newcolumntype{C}[1]{>{\hsize=#1\hsize\centering\arraybackslash}X}%

\newcommand{\Mearth}{\, M_{\oplus}}

\title[Hot Jupiters and inner super-Earths]{In situ formation of hot Jupiters with companion super-Earths}

\author[Poon, Nelson, and Coleman]{
Sanson T. S. Poon$^{1,2}$\thanks{E-mail: s.t.s.poon@qmul.ac.uk},
Richard P. Nelson$^{1}$, and
Gavin A. L. Coleman$^{1}$
\\
$^{1}$Astronomy Unit, Queen Mary University of London, London E1 4NS, UK\\
$^{2}$Royal Observatory Greenwich, London SE10 9NF, UK\\
}

\date{Accepted 2021 May 15. Received 2021 May 6; in original form 2020 December 24.}

\pubyear{2021}

\begin{document}
\label{firstpage}
\pagerange{\pageref{firstpage}--\pageref{lastpage}}
\maketitle

\begin{abstract}
Observations have confirmed the existence of multiple-planet systems containing a hot Jupiter and smaller planetary companions. Examples include WASP-47, Kepler-730, and TOI-1130. We examine the plausibility of forming such systems \textit{in situ} using $N$-body simulations that include a realistic treatment of collisions, an evolving protoplanetary disc and eccentricity/inclination damping of planetary embryos. Initial conditions are constructed using two different models for the core of the giant planet: a `seed-model' and an `equal-mass-model'. The former has a more massive protoplanet placed among multiple small embryos in a compact configuration. The latter consists only of equal-mass embryos. 
Simulations of the seed-model lead to the formation of systems containing a hot Jupiter and super-Earths. The evolution consistently follows four distinct phases: early giant impacts; runaway gas accretion onto the seed protoplanet; disc damping-dominated evolution of the embryos orbiting exterior to the giant; a late chaotic phase after dispersal of the gas disc. Approximately 1\% of the equal-mass simulations form a giant and follow the same four-phase evolution. Synthetic transit observations of the equal-mass simulations provide an occurrence rate of 0.26\% for systems containing a hot Jupiter and an inner super-Earth, similar to the 0.2\% occurrence rate from actual transit surveys, but simulated hot Jupiters are rarely detected as single transiting planets, in disagreement with observations. A subset of our simulations form two close-in giants, similar to the WASP-148 system. The scenario explored here provides a viable pathway for forming systems with unusual architectures, but does not apply to the majority of hot Jupiters.
\end{abstract}



\section{Introduction}
Since the discovery of the hot Jupiter 51 Pegasi b \citep{1995Natur.378..355M}, observations using the radial velocity (RV) technique have unveiled a large number of such objects
\citep[e.g.][]{2002A&A...390..267U,2006A&A...446..717D,2006ApJ...646..505B,2012ApJ...756L..33Q,
2019ApJ...878L..37F}.\footnote{All exoplanet data used in this paper are from the NASA Exoplanet Archive unless stated otherwise.} The population study by \citet{2011arXiv1109.2497M} indicates that $\sim 14\%$ of Sun-like stars host a gas giant planet with mass $\gtrsim 50 \Mearth$ and orbital period $\lesssim 5000$ days, and the fraction of stars hosting such planets with periods $\lesssim 10$ days is $\sim 1\%$. 

Various scenarios have been proposed to explain the origins of hot Jupiters, including formation at larger distances followed by migration driven by the gaseous protoplanetary disc, \textit{in situ} formation after the build-up of a large reservoir of planetary building blocks close to the star, and migration due to tidal interaction with the central star after a giant planet achieves a high eccentricity orbit due to planet-planet scattering, or through the Kozai-Lidov effect induced by a distant companion (see \citet{2018ARA&A..56..175D} and references therein for a comprehensive discussion on the origins of hot Jupiters).

Transit surveys have also discovered hot Jupiters. Of particular interest are the discoveries of multiple systems that contain a transiting hot Jupiter and inner planetary companions, such as WASP-47 \citep{2015ApJ...812L..18B}, Kepler-730 \citep{2019ApJ...870L..17C}, and TOI-1130 \citep{2020ApJ...892L...7H}. These multiple planet systems cannot have formed through high-eccentricity migration, and hence must have formed \textit{in situ} or after large scale migration. Previous studies have examined the consequences for planet formation of a giant planet migrating over large distances, and have shown that a natural outcome can be the formation of relatively compact multi-systems consisting of hot Jupiters and nearby super-Earths/mini-Neptunes \citep[e.g.][]{2005A&A...441..791F,2007A&A...461.1195F,2006Sci...313.1413R,2007ApJ...660..823M}. Our aim in this study is examine whether or not such systems naturally arise from an \textit{in situ} formation scenario, for which a key assumption is that disc-driven migration does not occur.

One argument deployed in favour of the migration scenario for the formation of hot Jupiters is that a protoplanetary disc with mass characteristic of the Minimum Mass Solar Nebula model \citep[MMSN,][]{1981PThPS..70...35H} lacks sufficient mass in solids and gas to build a planet \textit{in situ}. Instead, it is envisaged that planet cores form beyond the snow line and start to accrete their gas envelopes at large distances from the star \citep{1996Icar..124...62P}. Subsequent angular momentum exchange with the disc leads to inwards migration \citep[e.g.][]{1996Natur.380..606L,2002ApJ...565.1257T,2004MNRAS.350..849N,2005A&A...434..343A,
2012ARA&A..50..211K,2014MNRAS.445..479C,2016MNRAS.460.2779C,2019A&A...623A..88B}, during which accretion of gas onto the planet continues. Various arguments have been used against this scenario. For example, \citet{2018ApJ...866L...2B} suggest that the inner envelope of the mass-period distribution for hot Jupiters is best explained by a model in which hot Jupiters accrete their gas \textit{in situ}. This argument hinges on both the planet mass and the size of the magnetospheric cavity in a protoplanetary disc depending on the mass accretion rate through the disc. It is possible a migration based scenario might also produce such a relation if a migrating planet accretes gas through a gap at a rate that scales with the overall accretion rate through the disc. The difference between the planetary mass distributions when comparing hot and cold giants, where the mean mass of hot Jupiters is slightly lower than that of cold Jupiters, also seems to be in tension with the migration scenario, since this trend is not expected if hot and cold Jupiters both formed at large distance and hot Jupiters continue to accrete as they migrate.

Transit surveys such as Kepler have discovered a large population of close-in planets \citep{2010Sci...327..977B,2011ApJ...736...19B,2013ApJS..204...24B, 2014ApJS..210...19B,2015ApJS..217...16R,2015ApJS..217...31M, 2016ApJS..224...12C,2018ApJS..235...38T}, and have shown that systems of multiple super-Earths/mini-Neptunes with orbital period $<100$ days are common around Sun-like stars \citep[e.g.][]{2013ApJ...766...81F,2013ApJ...767...95D,2015ApJ...807...45D, 2013PNAS..11019273P,2018ApJ...860..101Z}. Some of these compact planetary systems contain a significant amount of mass \citep[e.g.][]{2011Natur.470...53L}, and a number of studies have been conducted to examine \textit{in situ} formation scenarios, where it is assumed a large amount of solid mass has drifted into the inner disc before accumulating into planetary building blocks which then collide to form compact planetary systems \citep[e.g.][]{2013MNRAS.431.3444C,2013ApJ...775...53H,2016ApJ...832...34M,2017AJ....154...27M,
2018MNRAS.478.2896M,2020MNRAS.491.5595P,2020A&A...642A..23M}. A natural extension of this scenario is to suppose that hot Jupiters can form \textit{in situ} if sufficiently massive cores form through collisional accretion during the gas disc lifetime such that they can accrete massive envelopes \citep[e.g.][]{2014ApJ...797...95L,2016ApJ...817L..17B,2016ApJ...829..114B,2019A&A...629L...1H}. 

A key assumption adopted by the above cited \textit{in situ} models is that gas disc-induced planetary migration can be neglected completely. Some justification for this comes from the fact that migration occurs because of the net torque that results from the competition between contributions from inner and outer Lindblad resonances and the corotation region, and these can depend in a complicated way of the local conditions in the protoplanetary disc that are not well understood. Both the direction and speed of migration depend on the local level of turbulent viscosity, the temperature gradient and the thermal relaxation timescale, among other disc properties. In order to examine the specific issue of what types of planetary system architectures arise from \textit{in situ} models of giant planet formation, we also adopt the assumption that migration can be neglected. We refer readers who are interested in comparing the \textit{in situ} models presented here with those that include migration to our previous studies \citep{2014MNRAS.445..479C,2016MNRAS.457.2480C,2016MNRAS.460.2779C}.

In this paper we use $N$-body simulations to study the \textit{in situ} formation of planetary systems containing a hot Jupiter and at least one inner planetary companion. We assume there is sufficient solid material in the inner region of the protoplanetary disc (a total of $\sim25$ to $30\,\mathrm{M_{\oplus}}$ at $\lesssim0.5\,\mathrm{au}$) to build a core that can undergo runaway gas accretion. The solid mass is distributed among multiple ($N>50$) protoplanets which dynamically evolve and collide to build the final planetary system, and if a planet grows to be large enough during the gas disc lifetime then it is able to accrete gas and become a giant planet. We adopt two sets of initial conditions, namely the `seed-model' and the `equal-mass-model'. The seed-model initially contains a more massive seed-protoplanet that ensures a hot giant can form in the simulation. This biased initial condition allows us to study the dynamical history and behaviour of the \textit{in situ} gas giant formation model. The equal-mass-model initially contains equal mass protoplanets, where numerous protoplanet mergers in the early stages of a simulation are essential to form a giant planet. This model allows us to investigate the occurrence rate of giant planets in the \textit{in situ} model in an unbiased manner.

The possibility of significant gravitational scattering of protoplanets during the growth of a giant planet, combined with the large Keplerian velocities close to the star, have motivated us to adopt a realistic collision algorithm in our $N$-body simulations \citep{2012ApJ...745...79L}. We employ a 1D evolving viscous disc model, which is used to implement eccentricity and inclination damping forces onto the protoplanets. Migration torques are neglected in this study to be consistent with the \textit{in situ} formation scenario. Our model includes gas accretion onto planetary cores when they become sufficiently massive, gap opening, and photoevaporation of the disc. The realistic collision model allows us to track the nature of planet-planet collisions, and the generation of collisional debris during high energy impacts, and we use this information to compare the collisional behaviour between the gas-rich and gas-free phases. The giant formation rate is sensitive to the gas envelope accretion model adopted, and we examine the effect of this on the outcomes of our simulations.

This paper is structured as follows. In Section \ref{sec:simmodel}, we describe the simulation methods, including the realistic collision model and the disc model. In Section \ref{sec:setup}, we describe the initial conditions. In Section \ref{sec:mainresult}, we present the results and analyses of our simulations. In Section \ref{sec:opa}, we investigated the impact of assuming different gas envelope accretion models. Finally, we discuss our results and draw conclusion in Section \ref{sec:discon}.

\section{Simulation model}\label{sec:simmodel}
We use the $N$-body code {\sc symba} \citep{1998AJ....116.2067D} to undertake the simulations presented in this paper, and we utilise a version that implements an algorithm to treat planet-planet collisions realistically. We also include an evolving protoplanetary disc model to provide a prescription for planet-disc interactions, which we describe below.

\subsection{Realistic collision model}
The \citet{2012ApJ...745...79L} collision model was implemented in {\sc symba} as described our previous study \citep{2020MNRAS.491.5595P}. Here we just summarise the post-collision outcomes that can arise in the simulations.
For a more detailed description we refer the reader to the above cited papers \citep[see also][]{2010ApJ...714L..21K,
2012ApJ...744..137G,
2012ApJ...751...32S,
2020MNRAS.493.4910S}.

We refer to the more massive body involved in a collision as the target, and the less massive object as the projectile. Post collision, we can have a largest remnant, a second largest remnant and debris in the form of superplanetesimals. Superplanetesimals are non-mutually interacting gravitating particles with masses between 0.1 to 10 times the mass of Ceres. The radii of all objects are calculated using the mass-radius relation equation \ref{eq:mass-radius}. The collision algorithm consists of a decision tree with the following possible outcomes that depend on how the impact velocity, $V_{\rm imp}$, compares with the escape velocity, $V_{\rm esc}$:
\begin{enumerate}
\item Perfect merger, when $V_{\mathrm{imp}}<V_{\mathrm{esc}}$. A single body is formed with the total mass and momentum of the original two collided bodies.
\item Supercatastrophic disruption, when $V_{\mathrm{imp}}$ exceeds the threshold for supercatastrophic disruption. No massive bodies remain and all the mass is in the form of collision debris represented by superplanetesimals.
\item Catastrophic disruption, when $V_{\mathrm{imp}}$ exceeds the threshold for catastrophic disruption. Only one massive body remains and the rest of the mass is in the form of collision debris.
\item Erosion, when $V_{\mathrm{imp}}$ exceeds the threshold for erosion. The target is eroded by the projectile and loses some of its mass, while the projectile is completely disrupted into superplanetesimals.
\item Partial accretion, when $V_{\mathrm{imp}}$ is less than the threshold for erosion and the collision angle is smaller than the critical angle. The target gains mass from the projectile, which is completely disrupted into numerous superplanetesimals.
\item Hit-and-spray, when the collision angle is larger than the critical collision angle and $V_{\mathrm{imp}}$ exceeds the velocity of transition to hit-and-spray. The projectile is completely disrupted into debris.
\item Hit-and-run, similar to the hit-and-spray criterion but with $V_{\mathrm{imp}}$ between the velocities for transition to hit-and-spray and hit-and-run. The projectile mass is reduced and the remaining mass goes into debris.
\item Bouncing collision, similar to the hit-and-run criterion but with $V_{\mathrm{imp}}$ smaller than the velocity for transition to hit-and-run and greater than the normalised critical impact velocity \citep{2010ApJ...714L..21K,2012ApJ...744..137G}. The target and projectile retain all of their mass and the collision is treated as an inelastic bounce.
\item Graze-and-merge, similar to bouncing criterion but with $V_{\mathrm{imp}}$ smaller than the normalised critical impact velocity. The outcome is similar to perfect merge where a single body forms containing all the mass of the colliding objects.
\end{enumerate}

When a supercatastrophic collision happens, it can lead to the formation of a stable ring composed of superplanetesimals that sits interior to the system of planetary embryos. This can be very stable and remain present throughout the simulations. To avoid integrating the large number of debris particles using a small time-step size, we adopted the mass reduction scheme described in \citet{2020MNRAS.491.5595P}, which is based on determining the collision frequency of the debris particles and assuming collisions will grind the debris to dust, which is then quickly removed from the system. 

\subsection{Protoplanetary disc model}\label{subsec:discmodel}
Here we describe the viscously evolving protoplanetary disc model and the eccentricity/inclination damping forces that are applied to the planetary embryos.

\subsubsection{Disc evolution}
The disc adopted here is a typical $\alpha$-disc model \citep{1973A&A....24..337S}. The kinematic viscosity, $\nu$, is given by
\begin{equation}\label{eq:viscosity}
\nu = \alpha c_{\mathrm{s}}^2\Omega^{-1} = \alpha \left ( \frac{k_{\mathrm{b}} T}{\mu m_{\mathrm{H}}}\right )\left ( \frac{GM_{\star}}{r^3} \right )^{-\frac{1}{2}},
\end{equation}
where $c_{\mathrm{s}}$ is the local sound speed, $k_{\mathrm{b}}$ is the Boltzmann constant, $\Omega$ is the local angular velocity, $T$ is the local temperature of the disc, $\mu$ is the mean molecular weight of the disc and set as $2.4\,\mathrm{u}$ in this study, $m_{\mathrm{H}}$ is the atomic mass of hydrogen, $G$ is the gravitational constant, $M_{\star}$ is the mass of the host star and $r$ is the local distance from the host star. We set $\alpha=10^{-3}$.

The evolution of the disc surface density, $\Sigma$, is given by the usual diffusion equation, augmented by additional terms that account for the torque exerted by a gap forming planet (if present in a simulation) and photoevaporation:
\begin{equation}\label{eq:Sigma}
\frac{\partial \Sigma}{\partial t}=\frac{1}{r}\frac{\partial}{\partial r} \left [ 3r^{1/2} \frac{\partial }{\partial r}\left ( \nu \Sigma r^{1/2} \right ) - \frac{2\Lambda \Sigma r^{3/2}}{\left (GM_{\star} \right )^{1/2}}  \right ]-\frac{\partial \Sigma_{\mathrm{pe}}}{\partial t},
\end{equation}
where $\Lambda$ is the injection rate of angular momentum per unit mass due to tidal interaction with a planet \citep{1986ApJ...309..846L}, and $\partial \Sigma_{\mathrm{pe}}/\partial t$ is the rate of change of the disc surface density due to photoevaporation (see equation \ref{eq:photoeva}). $\Lambda$ is given by
\begin{equation}\label{eq:Lambda}
\Lambda = \mathrm{sign}(r-r_{\mathrm{p}})\frac{GM_{\star}}{2r} \left ( \frac{M_{\mathrm{p}}}{M_{\star}} \right )^{2} \left ( \frac{r}{\max (H,|r-r_{\mathrm{p}}| )} \right )^{4},
\end{equation}
where $r_{\mathrm{p}}$ is the distance between the planet and the host star, $M_{\mathrm{p}}$ is the mass of the planet, and $H$ is the local disc scale height obtained from $H=c_{\mathrm{s}}/\Omega$. Equation \ref{eq:Lambda} is only applied when a planet mass reaches the gap opening criterion, otherwise $\Lambda = 0$. Gap opening occurs when the following criterion is satisfied \citep{2006Icar..181..587C}:
\begin{equation}\label{eq:gapopencri}
50\alpha\left ( \frac{M_{\star}}{M_{\mathrm{p}}} \right ) \left (\frac{H}{r_{\mathrm{p}}}  \right )^{2}+\frac{3}{4}\left ( \frac{3M_{\star}}{M_{\mathrm{p}}} \right )^{1/3}\left ( \frac{H}{r_{\mathrm{p}}} \right )-1 \leqslant 0.
\end{equation}

During the lates stage of a disc's lifetime, when the accretion rate is low, photoevaporation due to UV radiation from the star starts to dominate the disc surface density evolution \citep[e.g. ][]{2001MNRAS.328..485C,2003ApJ...582..893M,2004ApJ...605..880R,2005ApJ...627..286T}. The photoevaporation term, $\partial \Sigma_{\mathrm{pe}}/\partial t$ in equation \ref{eq:Sigma}, is obtained using the prescription in \citet{2007prpl.conf..555D}:
\begin{equation}\label{eq:photoeva}
\dfrac{\partial \Sigma_{\mathrm{pe}}}{\partial t}=1.16\times10^{-11} \Phi_{41} ^ {1/2} r_{\mathrm{g}}^{-3/2} S_{\mathrm{g}}\left ( \mathrm{\dfrac{M_{\odot}}{au^{2}\,yr}} \right ),
\end{equation}
where $r_{\mathrm{g}}$ is the gravitational radius, $S_{\mathrm{g}}$ is a scaling factor in terms of $r_{\mathrm{g}}$, and $\Phi_{41}$ is the rate of emitted ionizing photons from the host star in units of $10^{41}\,\mathrm{photons\,s^{-1}}$. The scaling factor $S_{\mathrm{g}}$ is calculated according to
\begin{equation}\label{eq:Define_S}
S_{\mathrm{g}}=\begin{cases}
\left ( \dfrac{r}{r_{\mathrm{g}}} \right )^{-2} \exp{\left ( \dfrac{r-r_{\mathrm{g}}}{2r}\right )} & \text{for~}r\le r_{\mathrm{g}},\\[10pt]
\left ( \dfrac{r}{r_{\mathrm{g}}} \right )^{-5/2} & \text{otherwise}.
\end{cases}
\end{equation}
The gravitational radius, $r_{\mathrm{g}}$ is a characteristic radius where 
the isothermal sound speed at the ionisation temperature of the gas is equal to the Keplerian orbital velocity \citep[e.g.][]{1983ApJ...271...70B,1994ApJ...428..654H}. Thus,
\begin{equation}\label{eq:rg}
r_{\mathrm{g}}=\dfrac{GM_{\star}\mu m_{\mathrm{H}}}{k_{\mathrm{b}}T_{0}},
\end{equation}
where $T_{0}$ is the ionisation temperature of the gas \citep[$\sim10^{4}\,\mathrm{K}$, e.g.][]{2013ApJ...773..155T}. For a gas disc around a Sun-like star ($M_{\star}=1\,\mathrm{M_{\odot}}$), with $\mu=1\,\mathrm{u}$ and $T_{0}=10^{4}\,\mathrm{K}$ ($\sim$ ionisation temperature of hydrogen), the gravitational radius has a value of $r_{\mathrm{g}}\approx 10.8\,\mathrm{au}$. For all models presented below we adopt the values $r_{\mathrm{g}}=10\,\mathrm{au}$ and $\Phi_{41}=1$.

It is worth noting here that $r_{\mathrm{g}}$ does not define the radius exterior to which gas is able to escape the system, since a hydrodynamic flow is established that places gas into the escaping photoevaporative wind at radii $r < r_{\mathrm{g}}$ \citep[see discussion in][]{2007prpl.conf..555D}. This effect is accounted for in the prescription given by equations~\ref{eq:photoeva} and \ref{eq:Define_S}. Our adoption of a fixed value of $r_{\mathrm{g}}$ for all models, even though the masses of the host stars differ between them, leads to a small discrepancy between the life times of the disc models and what they would have been if a more accurate value of $r_{\mathrm{g}}$ were used. We have undertaken a suite of tests to examine the effects of this on our simulation results and find that the differences are negligible.

\subsubsection{Eccentricity and inclination damping}
A planet orbiting in a gaseous disc will experience eccentricity and inclination damping forces. It may also experience torques that drive migration, but in this study we neglect these and examine how systems evolve only under the influence of eccentricity/inclination damping, while remaining agnostic about why migration does not occur. Migration arises from a competition between torques originating at inner and outer Lindblad resonances and in the corotation region, and both the direction and speed of migration can vary depending on local disc conditions. For an embedded planet it is expected that damping of eccentricity and inclination will always occur.

We follow \citet{2000MNRAS.315..823P} in our implementation of the damping forces. The accelerations of an object due to the eccentricity damping, $\pmb{a}_{\mathrm{e\text{-}damp}}$, and inclination damping, $\pmb{a}_{\mathrm{i\text{-}damp}}$, are given by 
\begin{equation}
\pmb{a}_{\mathrm{e\text{-}damp}}=-2\dfrac{v_{\mathrm{r}}}{t_{\mathrm{e}}} \hat{\pmb{r}},
\end{equation}
and
\begin{equation}
\pmb{a}_{\mathrm{i\text{-}damp}}=-2\dfrac{v_{\mathrm{z}}}{t_{\mathrm{i}}}\hat{\pmb{z}},
\end{equation}
where $v_{\mathrm{r}}$ is the radial velocity, $v_{\mathrm{z}}$ is the velocity in the $\pmb{z}$ direction (perpendicular to the disc plane), $t_{\mathrm{e}}$ is the eccentricity damping timescale, and $t_{\mathrm{i}}$ is the inclination damping timescale.

We use the expressions from \citet{2008A&A...482..677C} to calculate $t_{\mathrm{e}}$ and $t_{\mathrm{i}}$, giving
\begin{equation}\label{eq:tedamp}
t_{\mathrm{e}}=\dfrac{t_{\mathrm{wave}}}{0.780} \left [ 1-0.14\bigg ( \dfrac{e}{h} \bigg )^{2}+0.06\bigg ( \dfrac{e}{h} \bigg )^{3}+0.18\bigg ( \dfrac{e}{h} \bigg ) \bigg ( \dfrac{I}{h} \bigg )^{2} \right ],
\end{equation}
and
\begin{equation}\label{eq:tidamp}
t_{\mathrm{i}}=\frac{t_{\mathrm{wave}}}{0.544}\left [ 1-0.30\bigg ( \dfrac{I}{h} \bigg )^{2}+0.24\bigg ( \dfrac{I}{h} \bigg )^{3}+0.14\bigg ( \dfrac{I}{h} \bigg ) \bigg ( \dfrac{e}{h} \bigg )^{2} \right ],
\end{equation}
where $e$ is the eccentricity, $I$ is the inclination, $h$ is the ratio between the disc scale height and the local radius ($H/r$), and $t_{\mathrm{wave}}$ is the characteristic time of the orbital evolution. These expressions were calibrated against 3D simulations with surface density power law distributions $\Sigma \sim \Sigma_0 r^{-1/2}$ in \citet{2008A&A...482..677C}, similar to those that arise in the constant $\alpha$-disc models we consider in this work. We note, however, that the surface density profile may change significantly during the phase when the disc is being photoevaporated.

\citet{2004ApJ...602..388T} defined $t_{\mathrm{wave}}$ as
\begin{equation}
t_{\mathrm{wave}}=\Bigg( \frac{M_{\mathrm{p}}}{M_{\star}} \Bigg)^{-1} \Bigg ( \frac{\Sigma_{\mathrm{p}} a_{\mathrm{p}}^{2}}{M_{\star}} \Bigg )^{-1} \Bigg ( \frac{c_{\mathrm{s}}}{a_{\mathrm{p}}\Omega_{\mathrm{p}}} \Bigg )^{4} \Omega_{\mathrm{p}}^{-1},
\end{equation}
where $a_{\mathrm{p}}$ is the semi-major axis of the object, and the values of $\Sigma_{\mathrm{p}}$, $c_{\mathrm{s}}$, and $\Omega_{\mathrm{p}}$ are evaluated at $a_{\mathrm{p}}$. For an Earth-like planet ($M_{\mathrm{p}}=1\,\mathrm{M_{\oplus}}$) orbiting around a solar-type star ($M_{\star}=1\,\mathrm{M_{\odot}}$) in the MMSN that we adopted in our simulation model ($\mu=2.4\,\mathrm{u}$, equations \ref{eq:initsigma} and \ref{eq:inittemp}), the typical value for $t_{\mathrm{wave}}$ is given by $t_{\mathrm{wave}}\approx 327 (r_\mathrm{p}/1\,\mathrm{au})^{2}\,\mathrm{yr}$.

The damping timescales $t_{\mathrm{e}}$ and $t_{\mathrm{i}}$ obtained from equations \ref{eq:tedamp} and \ref{eq:tidamp}, are applied until a planet satisfies the gap opening criterion (equation \ref{eq:gapopencri}). The gaps formed by tidal torques in 1D disc models, as presented here, tend to be very deep and do not agree with the results from hydrodynamic simulations of disc-planet interactions \citep[e.g.][]{2017MNRAS.469.3813H}. Hence, determining eccentricity and damping timescales based on the density of gas in the gap would be inaccurate. In this work we set the damping timescales to be 20 local orbital periods for gap opening planets, and for reference a Jovian mass planet orbiting at $0.05\,\mathrm{au}$ in a disc with mass three times larger than the MMSN, but located in a gap with a density contrast of 1\%, would experience a damping timescale of this magnitude. We note that the damping is removed once the gas disc is dispersed, and mutual dynamical interactions between the planets are then able to raise the eccentricities and inclinations. The observed systems we have used as templates either contain giant planets for which there are no reported measurements of the eccentricity (i.e.\,Kepler-730b, Kepler-487b) or have very low eccentricities, that are essentially consistent with zero within the error bars (i.e.\,WASP-47b has $e = 0.0028^{+0.0042}_{-0.0020}$ \citep{2016A&A...595L...5A}, Kepler-89d has $e=0.022 \pm 0.038$, TOI-1130c has $e=0.047^{+0.040}_{-0.027}$). These systems are consistent with evolution that included efficient damping of eccentricity and the absence of strong scattering of the giants, as displayed by the simulations presented here, although it should be noted that tidal interactions with the central star over Gyr timescales would have likely erased any significant eccentricities the observed giant planets may have been born with.

\subsection{Gas envelope accretion}\label{subsec:gasacc}
We allow gas to accrete onto protoplanets if their masses reach a threshold value. We adopt different gas accretion prescriptions to examine the sensitivity of our results to the assumed gas accretion rates, and these prescriptions are described below. For the main part of our study, we adopt a simple prescription for the gas accretion rate that is similar to or the same as one we have used in previous studies of planet formation \citep{2014MNRAS.445..479C,2016MNRAS.457.2480C,2016MNRAS.460.2779C}. This allows for direct comparison between the present work and these earlier studies. In order to examine the impact of adopting a different gas accretion prescription, we introduce a new and significantly more sophisticated approach based on a large suite of 1D envelope accretion models that are sensitive to local disc conditions. While these latter models are more realistic, they are also dependent on assumptions such as the opacity of the envelope, and hence cannot be compared directly with previous work.
\subsubsection{Simple model}\label{subsubsec:simplegasacc}
In our simple model, a protoplanet can start to accrete a gaseous envelope from the protoplanetary disc when its mass reaches $3\,\mathrm{M}_{\oplus}$. We adopt the expression for the gas accretion rate from \citet{2016MNRAS.457.2480C}, which provides a fit to the 1D calculations presented in \citet{2010Icar..209..616M} for a planet located at 5.2\,au from its host star:
\begin{equation}\label{eq:accretionrate}
\dfrac{d M_{\mathrm{ge}}}{d t}=\dot{M}_{\mathrm{0}} \left (\dfrac{M_{\mathrm{core}}}{\mathrm{M}_{\oplus}}  \right )^{2.4} \exp{\left [\dfrac{M_{\mathrm{ge}}}{\mathrm{M}_{\oplus}} \left (\dfrac{1}{22} + \dfrac{\mathrm{M}_{\oplus}}{M_{\mathrm{core}}}  \right )  \right ]}.
\end{equation}
Here $\dot{M}_{\mathrm{0}}$ is the mass accretion rate for a $1\,\mathrm{M_{\oplus}}$ protoplanet core (with no gas envelope), $M_{\mathrm{core}}$ is the mass of the protoplanet core, $M_{\mathrm{ge}}$ is the mass of the accreted gas envelope, and we have the relation $M_{\mathrm{p}}=M_{\mathrm{core}}+M_{\mathrm{ge}}$. The value of $\dot{M}_{\mathrm{0}}$ used is that in \citet{2016MNRAS.457.2480C}, namely $\dot{M}_{\mathrm{0}}=4.656\times10^{-8}\,\mathrm{M_{\oplus}\,yr^{-1}}$. The simple model, equation \ref{eq:accretionrate}, is adopted in all the simulations presented in section \ref{sec:mainresult}.

\subsubsection{Gas accretion based on fits to local 1D models}\label{subsubsec:newgasacc}
We also run a subset of simulations using a newer local gas accretion model (see section \ref{sec:opa}). For this model, we calculated new fits to gas accretion rates obtained using the 1D envelope structure model of \citet{2017MNRAS.470.3206C} \citep[see also][]{1999ApJ...521..823P,2005A&A...433..247P}. Whilst the gas accretion rates used in \citet{2016MNRAS.457.2480C} were based on fits from a handful of simulations in \citet{2010Icar..209..616M}, they do not take into account the local disc properties, i.e.\,surface density and temperature. This was appropriate for those models since the simulations performed in \citet{2010Icar..209..616M} were based on formation scenarios for Jupiter located at $5.2\,\mathrm{au}$.

In taking the local disc properties into account, we ran a large number of simulations where we placed planetary cores of masses between $2-15\,\mathrm{M}_{\oplus}$ at orbital radii spanning $0.2-50\,\mathrm{au}$ in the evolving disc model of \citet{2016MNRAS.457.2480C}. Using the gas envelope and accretion model of \citet{2017MNRAS.470.3206C}, these cores were able to accrete gas until the protoplanetary disc dispersed, or until they reached a critical state where runaway gas accretion occurs and they become giant planets. We included an opacity reduction factor $f_{\mathrm{opa}}$ that reduces the grain opacity contribution, similar to other works \citep[e.g.][]{2014A&A...566A.141M}.

As the discs were evolving, the local disc properties for the planets were ever-changing and this was taken into account when calculating the 1D gas envelope structure. These local disc properties were recorded, as well as the planet properties (e.g. core mass, envelope mass) and the mass accretion rate. In total the simulations provided $\sim 50,000$ data points to determine a gas accretion rate, a significant improvement on the 20 data points used to formulate the equations in \citet{2016MNRAS.457.2480C}. With these results we were able to fit an equation that takes into account not only the properties of the protoplanets, but also the local properties of the disc, notably the temperature, $T_{\mathrm{local}}$, and the opacity reduction factor, $f_{\mathrm{opa}}$. The local gas accretion rate can then be calculated according to
\begin{align}\label{eq:gasenvelope-gc}
\left (\dfrac{d M_{\mathrm{ge}}}{d t} \right )_{\mathrm{local}}=& 10^{-10.199} \left( \dfrac{\mathrm{M_{\oplus}}}{\mathrm{yr}}\right) f_{\mathrm{opa}}^{-0.963} \left ( \dfrac{T_{\mathrm{local}}}{\mathrm{1\,K}}\right )^{-0.7049}\nonumber \\
&\times \left (\dfrac{M_{\mathrm{core}}}{\mathrm{M}_{\oplus}} \right )^{5.6549}  \left (\dfrac{M_{\mathrm{ge}}}{\mathrm{M}_{\oplus}}  \right )^{-1.159}\nonumber \\
&\times\left [ \exp{\left ( \dfrac{M_{\mathrm{ge}}}{M_{\mathrm{core}}} \right )} \right ]^{3.6334}.
\end{align}
The strong dependence on the core mass in equation \ref{eq:gasenvelope-gc} is notable, as is the dependence of the gas envelope mass fraction. Such strong dependences are in agreement with \citet{2017MNRAS.470.3206C} where the accretion rate is heavily dependent on the core and envelope masses. The dependence on the temperature is also consistent with previous works where gas accretion rates decrease as the local disc temperature rises. Note that the effect of surface density is not included in this equations, since it is found to have a very weak effect. 
\begin{figure}
\begin{center}
\includegraphics[width=\columnwidth]{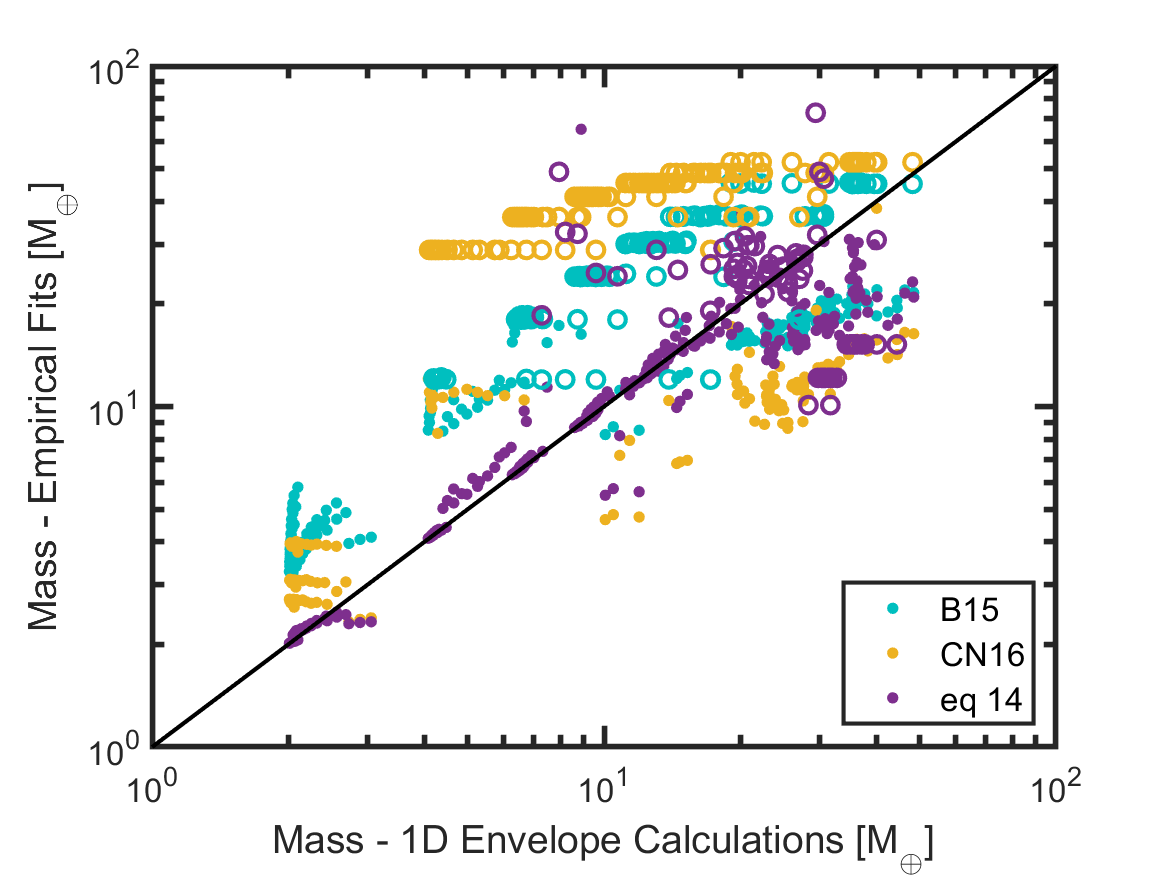}
\caption{Comparison of final planet masses from gas accretion fits to those calculated using the 1D gas envelope structure model of \citet{2017MNRAS.470.3206C}. The different colour markers represent different accretion fits arising from (cyan) \citet[][B15, their equation 17]{2015A&A...582A.112B}, (yellow) \citet[][CN16]{2016MNRAS.457.2480C}, and (purple) equation~\ref{eq:gasenvelope-gc}. The open circle markers indicate where the planets have reached the runaway gas accretion phase, and would become giant planets. The black line indicates a 1:1 relation between the final planet masses obtained using the fits and the 1D model.}
\label{fig:new_gas_rate}
\end{center}
\end{figure}

When comparing the planet masses attained using equation \ref{eq:gasenvelope-gc} to those arising from the 1D envelope structure model, we find that they are in good agreement. Figure \ref{fig:new_gas_rate} shows the final planet masses arising from calculations using the 1D envelope structure model of \citet{2017MNRAS.470.3206C} ($x$-axis) versus those attained from different gas accretion fits ($y$-axis). The purple points correspond to gas accretion fits based on the 1D envelope model (equation \ref{eq:gasenvelope-gc}), the yellow points come from the fits in \citet{2016MNRAS.457.2480C} to the work of \citet{2010Icar..209..616M},
and the cyan points arise from the fits in \citet[][B15]{2015A&A...582A.112B} to the envelope accretion calculations in \citet{2014ApJ...786...21P}.

To calculate these masses, we ran numerous single-planet-in-a-disc models where we placed cores of different masses at different locations in the disc, and allowed them to accrete locally. We did not include any migration in these models, and we stopped each model when the planets reached either: a critical state where the planet envelope can no longer be hydrostatically supported (1D models), a gas accretion rate of $2\,\mathrm{M_{\oplus}}/{1000\,\mathrm{yr}}$ (CN16 model, equation \ref{eq:gasenvelope-gc}) and where the envelope mass was equal to double the core mass (the B15 model). We denote planets that reach a critical state using open circles, whereas closed circles denote planets that did not reach the relevant criterion. In all of these scenarios, a period of runaway gas accretion is expected to occur with the planets becoming giant planets.

With the final masses, we then compare those masses calculated using the gas accretion rates based on empirical fits ($y$-axis) to the those masses calculated using the 1D envelope structure model ($x$-axis). The final planet masses obtained using equation \ref{eq:gasenvelope-gc} (purple markers) sit close to the black diagonal line that represents a 1:1 ratio between the planet masses found through the 1D envelope structure model and the gas accretion fits. This is especially true for low mass planets, i.e. super-Earth mass range. The only region where the fits become less consistent with the 1D model is at higher planet masses, where the fits reach a critical state slightly earlier or later than occurs in the 1D model. This will result in only small differences in the final planet masses, since these occurrences happen early in the disc lifetime, and the planets in all cases will undergo runaway gas accretion and become giant planets.

When comparing the results from other works to the 1D structure models, for the higher mass planets it is clear that like the masses found through equation~\ref{eq:gasenvelope-gc}, the planets reached a critical state slightly before/after the 1D model, where again these planets would undergo runaway gas accretion and become giant planets. Where the results from other works disagree with the 1D model is at lower planet masses particularly between $4\,\mathrm{M}_{\oplus}\leq M_{\mathrm{p}} \leq 15\,\mathrm{M}_{\oplus}$. For both accretion fits (CN16 and B15), the planets typically reached a critical state, whereas the planets in the 1D models did not. This would ultimately lead to the CN16 and B15 fits producing an over abundance of giant planets compared to the 1D model results.

For much lower mass planets, $M_{\mathrm{p}}\le 4\,\mathrm{M}_{\oplus}$, the accretion rates from \citet{2015A&A...582A.112B} and \citet{2016MNRAS.457.2480C} do not reach a critical state. This is more consistent with the 1D models, where the planets do not have large enough core masses to be able to accrete significant gaseous envelopes. However, even though there is greater agreement between the 1D models and the fits here, the masses arising from the accretion rate fits are still considerably larger than those arising from the 1D models. This is in contrast to the masses arising from equation~\ref{eq:gasenvelope-gc} where there is excellent agreement with the 1D models at low planet mass.

\subsubsection{Accretion after gap opening}
The adopted gas accretion model (equation~\ref{eq:accretionrate} or \ref{eq:gasenvelope-gc}) applies until a planet satisfies the gap opening criterion (equation~\ref{eq:gapopencri}) or when a protoplanet is sitting in a gap which was opened by another massive protoplanet. The accretion rate then changes to
\begin{equation}\label{eq:accdomain}
\left(\frac{d M_{\mathrm{ge}}}{d t}\right)_{\mathrm{gap}}=\min\left [\dfrac{3\upi\nu_{\mathrm{vsd}}\Sigma_{\mathrm{vsd}}}{N_{\mathrm{p,gap}}},\left(\frac{d M_{\mathrm{ge}}}{d t}\right)_{\mathrm{local}}  \right ],
\end{equation}
where $N_{\mathrm{p,gap}}$ is the number of protoplanets that sit in the gap and are accreting gas through viscous supply, $\nu_{\mathrm{vsd}}$ is the local viscosity and $\Sigma_{\mathrm{vsd}}$ is surface density in the disc exterior to the gap. To ensure that the numerator $3 \upi\nu_{\mathrm{vsd}}\Sigma_{\mathrm{vsd}}$ measures the typical accretion rate through the disc, and not the local value in the gap, we evaluate it at 2\,au from the star.

\begin{figure}
\begin{center}
\includegraphics[width=\columnwidth]{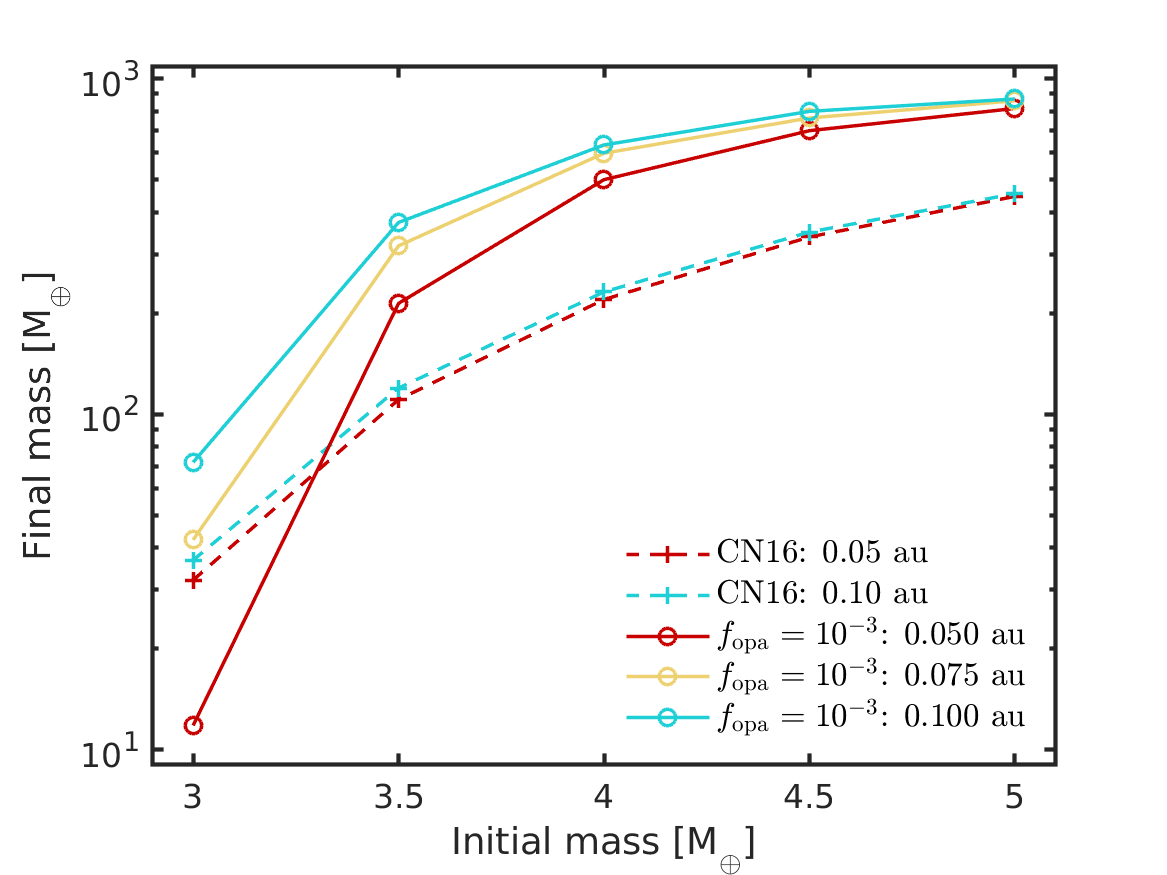}
 \caption[Initial and final planetary masses of the gas accretion routines]{Comparison between the initial and final masses (at the point of gas disc dissipation) of the planet in the single-planet simulations, the difference is based on the adoption of different gas envelope accretion models and initial orbital radii. The results of applying the \citet[][CN16]{2016MNRAS.457.2480C} routine (equation \ref{eq:accretionrate}) are plotted using dashed lines, where the protoplanet is initially located at 0.05\,au (red) or 0.10\,au (cyan). The solid lines represent the results obtained when adopting our locally-dependent envelope accretion routine (equation \ref{eq:gasenvelope-gc}) with the value of $f_{\mathrm{opa}}=10^{-3}$, where the protoplanet is initially located at 0.05\,au (red), 0.075\,au (yellow), or 0.10\,au (cyan). It is clear that the CN16 routine is less dependent on the initial protoplanet mass and orbital location (local disc conditions).
 }\label{fig:mass-init-fin}
\end{center}
\end{figure}

In figure \ref{fig:mass-init-fin} we demonstrate the difference between applying equation~\ref{eq:accretionrate} and \ref{eq:gasenvelope-gc} when calculating the gas accretion rate, including the switch to equation~\ref{eq:accdomain} once a gap has been opened. The final masses obtained when adopting equation~\ref{eq:gasenvelope-gc} shows a higher sensitivity to the local disc parameters compared to the results from equation~\ref{eq:accretionrate}. This mainly arises due to the increase in disc midplane temperature closer to the star, reducing the gas accretion onto the planets and having a much larger effect for smaller core masses.

For the main simulation results and analysis presented in section~\ref{sec:mainresult}, we use the accretion rates provided by equations~\ref{eq:accretionrate}, while in section~\ref{sec:opa} we apply equation~\ref{eq:gasenvelope-gc} for comparison.

\section{Simulation set-up}\label{sec:setup}
\begin{figure}
\begin{center}
\includegraphics[width=1.0\columnwidth]{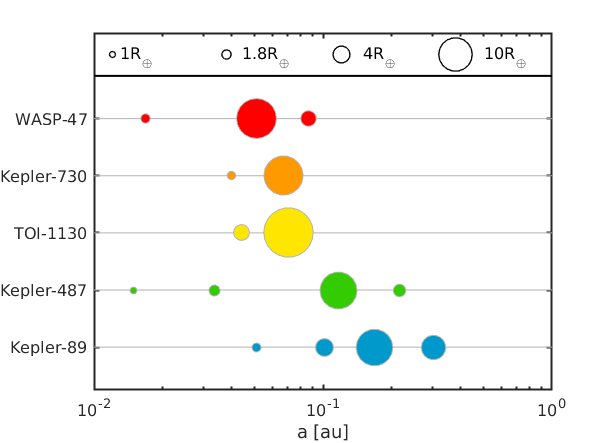}
\caption[Selected systems with a hot Jupiter and inner transiting companions]{The five selected planetary systems that meet the selection criteria (section \ref{sec:setup}). The symbol size represents the radius of each planet.
}\label{fig:selectedsystem}
\end{center}
\end{figure}

In this study, we are interested in systems of hot Jupiters with coexisting inner super-Earths. We have selected a number of systems that have a hot Jupiter and an inner companion to provide templates for the initial conditions of the simulations. The consideration of system selection is straight forward and uses the following criteria: (i) the system contains a confirmed transiting giant with orbital period less than 30 days; (ii) the system contains at least one transiting companion with orbital period shorter than the orbital period of the giant. There are five systems that meet these criteria. In order of increasing orbital period of the giant they are WASP-47, Kepler-730, TOI-1130, Kepler-487\footnote{The two inner companions, KOI-191.02 and KOI-191.03, of Kepler-487 are candidates instead of confirmed planets. We include this system because both objects have a relatively low probability to be due to any of the considered astrophysical false positive scenarios ($1.4\times10^{-3}$ for KOI-191.02 and $4.2\times10^{-5}$ for KOI-191.03).}, and Kepler-89 (see {figure \ref{fig:selectedsystem}}, and also table \ref{tab:systemparameter} for basic stellar and planetary parameters of the selected systems). 

\subsection{Template construction}
We constructed two sets of templates for our simulations: the `seed-model' and the `equal-mass model'. The seed-model contains one higher mass protoplanet, that acts as a seed for the growth of a giant planet, and multiple equal-mass smaller bodies. The equal-mass-model only contains planetary embryos of equal mass. The configurations for the two templates are:
\begin{enumerate}
\item seed-model: 25 inner equal-mass bodies + 1 seed protoplanet + 25 outer equal-mass bodies,
\item equal-mass-model: 51 equal-mass bodies.
\end{enumerate}
All the equal-mass bodies have mass $0.5\,\mathrm{M_{\oplus}}$. There are two subsets of the seed-model, where the mass of the seed protoplanet is either $4.0$ or $4.5\,\mathrm{M_{\oplus}}$. The middle body for each template (the $26\mathrm{th}$ body) is located at the semi-major axis of the hot Jupiter from the selected system. The mutual separations between each adjacent body are $K=5$ for the seed-model, and two subsets of $K=4$ and $5$ for the equal-mass model. This $K$-value is measured in units of the mutual Hill radius, $R_{\mathrm{H}}$, which is defined by
\begin{equation}\label{eq:Ki}
K=\frac{a_{i+1}-a_{i}}{R_{\mathrm{H},i}},
\end{equation}
where
\begin{equation}\label{eq:R_Hi}
R_{{\mathrm{H}},i}=\frac{a_{i}+a_{i+1}}{2}\left ( \frac{M_{\mathrm{p},i}+M_{\mathrm{p},i+1}}{3M_{\star}} \right )^{\frac{1}{3}},
\end{equation} 
and subscript $i$ denotes the value of the $i$-th body in the system. 

As discussed in section~\ref{subsec:discparameters}, the inner edge of the gaseous protoplanetary disc is located at $r=0.05\,\mathrm{au}$ for all models, and because we have centred the radial distribution of embryos on the location of the giant planet for all templates, some of the embryos are located interior to the disc edge at the beginning of the simulations. The inner edge of the disc is presumed to arise because it is truncated by the stellar magnetosphere, and the location where this truncation occurs depends on the mass accretion rate through the disc. In this work we suppose the disc accretion rate was significantly larger during earlier stages of evolution prior to the simulations being initiated, such that the disc inner edge was closer to the star. It is assumed the embryos located close to the star were able to form in the disc during this earlier phase.

We use a labelling convention based on the parameters mentioned above when describing the simulations as follows: `(selected system)-(subset)'. In total, there are 20 different simulation templates from the combinations of 5 selected systems (\texttt{WASP47}, \texttt{Kepler730}, \texttt{TOI1130}, \texttt{Kepler487}, and \texttt{Kepler89}) and 4 different subsets (\texttt{4M-5K}, \texttt{4.5M-5K}, \texttt{XM-4K}, and \texttt{XM-5K}). 

For example, the \texttt{WASP47-4M-5K} template refers to the seed-model runs with a $4\,\mathrm{M_{\oplus}}$ seed-protoplanet (which is located at 0.0513\,au) and mutual separation $K=5$ for all pairs of bodies orbiting around a central star with $M_{\star}=1.040\,\mathrm{M_{\odot}}$, and \texttt{Kepler730-XM-4K} refers to the equal-mass model runs where the $26\mathrm{th}$ body is located at 0.0694\,au with a mutual separation $K=4$ for all pairs of embryos orbiting around a central star with $M_{\star}=1.047\,\mathrm{M_{\odot}}$. 

The differences between each of the systems are the stellar mass, stellar radius, and the location of the initial protoplanets. The central stars of each system have their masses and radii taken from table \ref{tab:systemparameter}. Likewise, the locations of the middle body (the $26\mathrm{th}$ body) of each system are taken from the semimajor axis of the giant stated in table \ref{tab:systemparameter}. The total solid mass is the same for all of the observed systems, but different for each sub-template because of the presence or otherwise of a more massive seed. All \texttt{XM} templates have a total solid mass of $25.5\,\mathrm{M_{\oplus}}$, while the \texttt{4M} and \texttt{4.5M} templates have $29\,\mathrm{M_{\oplus}}$ and $29.5\,\mathrm{M_{\oplus}}$ of solids respectively.

Each template is run with 10 different instances of the initial conditions. The initial eccentricities of the bodies are randomly drawn from a Rayleigh distribution with eccentricity parameter, $\sigma_{\mathrm{e}}=2\times10^{-3}$. The initial inclinations for each run are randomly drawn from a Rayleigh distribution with inclination parameter, $\sigma_{\mathrm{I}}=1\times10^{-3}\,\text{rad}$. The distributions follow the relation $e=2I$, but the initial values of $e$ and $I$ for each planet are independent. The arguments of pericentre, $\omega$, longitudes of ascending node, $\Omega$, and mean anomalies, $M$ are distributed uniformly in the range $0\leq(\omega,\Omega,M)<2\upi$. Objects whose orbital distances exceed 100\,au are removed from the simulations. The time steps used are set to be 1/20th of the shortest orbital period in the system. Each simulation is run for $10^7\,\mathrm{yr}$.

\subsection{Disc parameters}\label{subsec:discparameters}
The protoplanetary disc model (section \ref{subsec:discmodel}) is included in all $N$-body simulations. The disc surface density profile, $\Sigma_{\mathrm{init}}\left ( r \right )$, has the same power-law index as the MMSN \citep{1981PThPS..70...35H}:
\begin{equation}\label{eq:initsigma}
\Sigma_{\mathrm{init}}\left ( r \right ) = \Sigma_{\mathrm{init}}\left ( \mathrm{1\,au} \right )r^{-1.5},
\end{equation}
where $\Sigma_{\mathrm{init}}\left ( \mathrm{1\, au} \right )=1700\,\mathrm{g\,cm^{-2}}$ is the initial surface density of the disc at 1\,au. The temperature profile is fixed throughout the simulations and also follows the power index of the MMSN and is given by
\begin{equation}\label{eq:inittemp}
T\left ( r \right ) = T\left ( \mathrm{1\, au} \right )r^{-0.5},
\end{equation}
where $T\left ( \mathrm{1\,au} \right )=280\,\mathrm{K}$ is the temperature at 1\,au. Identical initial surface density and temperature profiles are adopted in all simulations. Although the initial disc profile is the same throughout our simulations, the evolution of the disc is not identical between each system due to the difference between the stellar masses.

For the main simulations (section \ref{sec:mainresult}), the gas envelope accretion routine uses equation \ref{eq:accretionrate}. For the investigation of the impact from the gas envelope accretion routine (section \ref{sec:opa}), the gas envelope accretion rate is given by equation \ref{eq:gasenvelope-gc}.

The inner and outer boundaries of the disc are located at $0.05$ and $30.0\,\mathrm{au}$, respectively. We apply a zero radial velocity condition at the outer boundary, which ensures that no additional mass flows into the disc model. We apply a zero torque condition at the inner boundary to allow accretion onto the star. Any bodies sitting outside of the boundaries of the disc do not interact with it, and so experience no eccentricity/inclination damping and do not accrete gas.

\section{Results}\label{sec:mainresult}
We now present the results of simulations that explore the \textit{in situ} formation of systems containing a hot Jupiter and inner super-Earths. All the gas envelope accretion calculations presented in this section \ref{sec:mainresult} are preformed by the simple model (section \ref{subsubsec:simplegasacc}, equation \ref{eq:accretionrate}).

\subsection{Evolution of the seed-model}
To recap, two sets of simulations were performed for the seed-model, one with the seed protoplanet mass equal to $4.0\,\mathrm{M_{\oplus}}$ (\texttt{4M-5K}), and the other with $4.5\,\mathrm{M_{\oplus}}$ (\texttt{4.5M-5K}). 

\subsubsection{Dynamical evolution}\label{subsec:seedevo}
The dynamical evolution for all the seed-models can be classified in to four distinct phases in terms of their behaviour. For convenience when discussing the different phases, we will refer to them as the \emph{early impact}, \emph{runaway gas accretion}, \emph{outer disc damping}, and \emph{late chaotic phases}. 

\begin{figure}
\begin{center}
\includegraphics[width=\columnwidth]{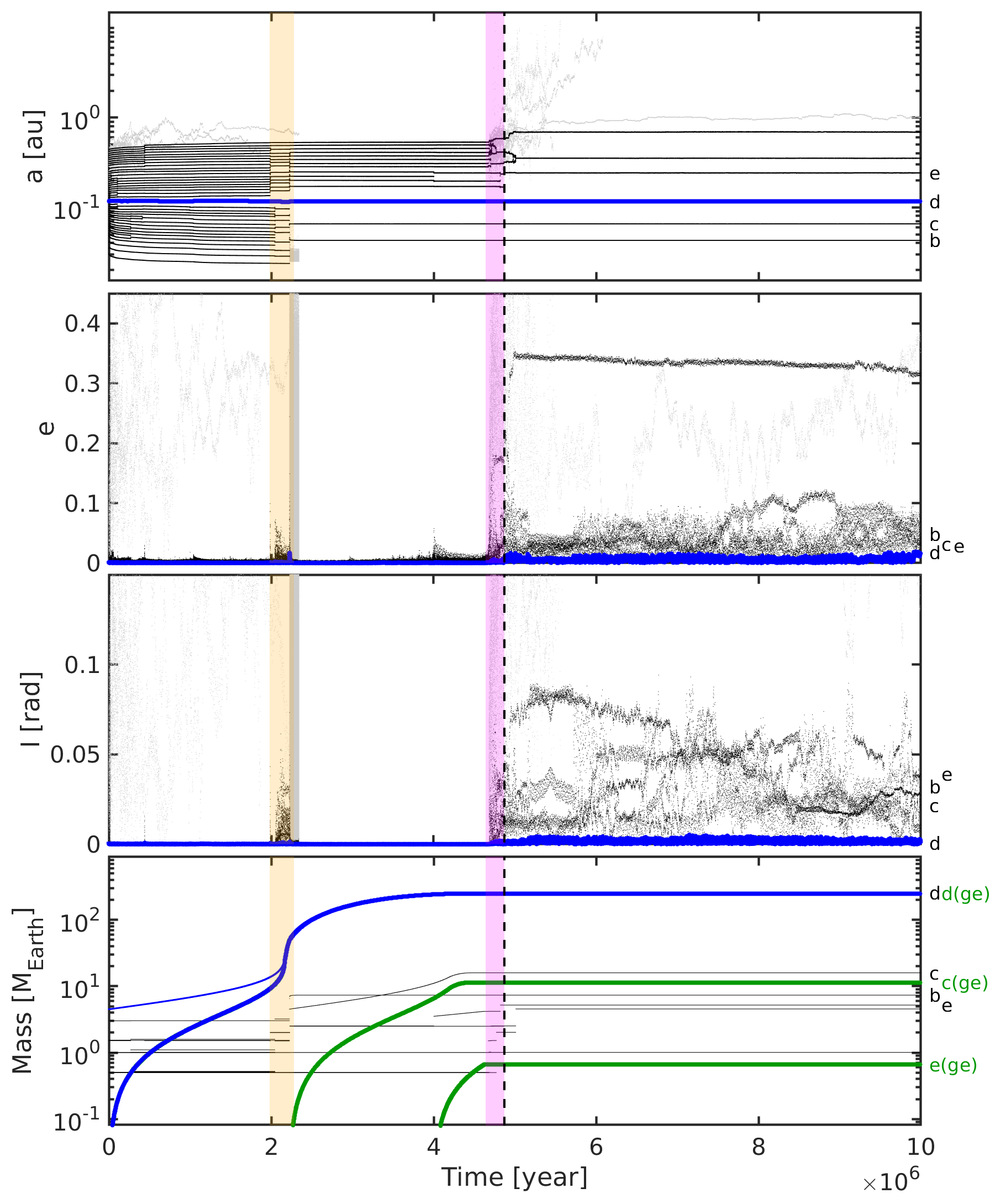}
 \caption[Dynamical history of the seed-model - \texttt{Kepelr487-4.5M}]{Dynamical history of a run from the \texttt{Kepler487-4.5M-5K} template, including the semimajor axes (top panel), eccentricities (second panel), inclinations (third panel), and planet masses (bottom panel). The seed-protoplanet is marked in blue, other protoplanets are marked in black, and collision debris are marked in grey. For the bottom panel, the thick lines are the gas envelope masses of the bodies. The gas envelope mass of the seed-protoplanet is marked in blue. The gas envelope masses of other protoplanets are marked in green. The black dashed vertical line denotes the point of disc dissipation. The orange shaded area marks the beginning of the runaway gas accretion phase. The magenta shaded area denotes the transition between the end of outer disc damping phase and the beginning of the late chaotic phase. The main planets formed (with final total planet mass $>5\,\mathrm{M_{\oplus}}$) are labelled as planet b, c, d, and e according to their final distance from the star, where planet d is the hot-Jupiter and planet b, c, and e are the companion super-Earths. The same set of planet labels is used in figure~\ref{fig:1d-disc}, which shows some instances of the same simulation. The green labels at the bottom panel indicate the final masses of the planet gas envelope (ge).}
 \label{fig:dym_evo_aeim}
\end{center}
\end{figure}

The early impact phase corresponds to the time early in the simulations when the embryos undergo dynamical instability and collisional accretion, before any planet in the system is massive enough to trigger its runaway gas envelope accretion (figure \ref{fig:dym_evo_aeim}, the first $\sim2\,\mathrm{Myr}$). The collisions between protoplanets in this stage are induced by the small initial mutual separations, and generally take place early in the simulations ($<1\,\mathrm{Myr}$ for initial $K=5$). After the initial impacts have occurred, the disc damping forces start to dominate the orbital evolution, and the eccentricities and inclinations are forced to remain very low. Except for the seed body, it is difficult to form a planet with mass greater than $4\,\mathrm{M_{\oplus}}$ because the damping forces prevent the occurrence of mergers.

As the name suggests, the runaway gas accretion phase starts when a planet enters runaway gas accretion (figure \ref{fig:dym_evo_aeim}, orange shaded area). The planet increases its mass exponentially in this stage and dynamically heats up the system. The dynamical heating by the swift increase in mass outstrips the disc damping forces, and orbit crossing of the bodies is a common outcome. This phase normally lasts for only a few hundred thousand years. Once the body enters the runaway accretion phase, especially in the inner parts of the discs that we consider in this study, it can open a gap and accrete gas at the viscous supply rate (equation~\ref{eq:accdomain}). This slows down the growth rate of the seed protoplanet and ends the runaway gas accretion phase.

The outer disc damping phase ensues after gap opening in the gas disc, and is illustrated in figure~\ref{fig:dym_evo_aeim} by the temporal domain lying between the orange and magenta shaded regions. In the parameter space that we consider in the seed-model, the outer edge of the planet-induced gap sits at an orbital radius $<0.15\,\mathrm{au}$. The inner part of the disc (in terms of the gap location) has a much shorter collisional accretion time compared to the outer part of the disc (disc beyond the gap). This phase is dynamically cooler than the previous runaway phase. The bodies in the inner part of the disc experience almost no disc damping forces, but the impacts experienced during the runaway phase result in relatively large mutual separations developing between the inner protoplanet pairs, so the inner system can be relatively stable. Inner protoplanets can grow to masses comparable to the initial seed mass, and start to accrete their own gas envelopes. However, these large cores initiate gas envelope accretion too late to become gas giants before the gas disc dissipates. The disc damping forces are still a dominating influence in the outer part of the disc. Protoplanets mergers are not very common during this phase, and generally we do not find that multiple massive cores form during the disc lifetime in the region sitting exterior to the giant planet's orbital radius.

The evolution enters the late chaotic phase when photoevaporation dominates the gas surface density evolution. In this stage, the density of the disc is low, and the damping forces gradually become negligible, until the disc eventually dissipates (figure \ref{fig:dym_evo_aeim}, magenta area and onward). Self-scattering between protoplanets can heat up the system dynamically, causing dynamical instabilities and giant impacts. Protoplanets can only accrete through giant impacts in this gas-free environment. This stage lasts until a long-term stable system emerges.

Throughout the evolution, the seed protoplanet retains a low eccentricity and inclination (figure \ref{fig:dym_evo_aeim}, blue line). The seed also essentially preserves its initial semimajor axis, and generally experiences just a $\sim 1\%$ decrease. For example, in figure \ref{fig:dym_evo_aeim}, the initial semimajor axis is 0.117\,au and its final value is 0.116\,au. This is due to exchanges of energy and angular momentum with the surrounding bodies, combined with the disc damping forces. 

We noticed that our seed for the \texttt{WASP47-4M} and \texttt{WASP47-4.5M} templates did not produce a hot giant in any of the runs for the above reason (the $\sim 1\%$ decrease of final semimajor axis comparing to the initial semimajor axis). The initial semimajor axes (0.051\,au) of these seeds are located very close to the inner edge of the disc (0.050\,au), and they stop accreting gas once the seeds move out of the disc. To form a hot giant orbiting with a semimajor axis like WASP-47b might require a seed initially at $\sim0.055\,\mathrm{au}$ for our model set-up.

\begin{figure}
\begin{center}
\includegraphics[width=\columnwidth]{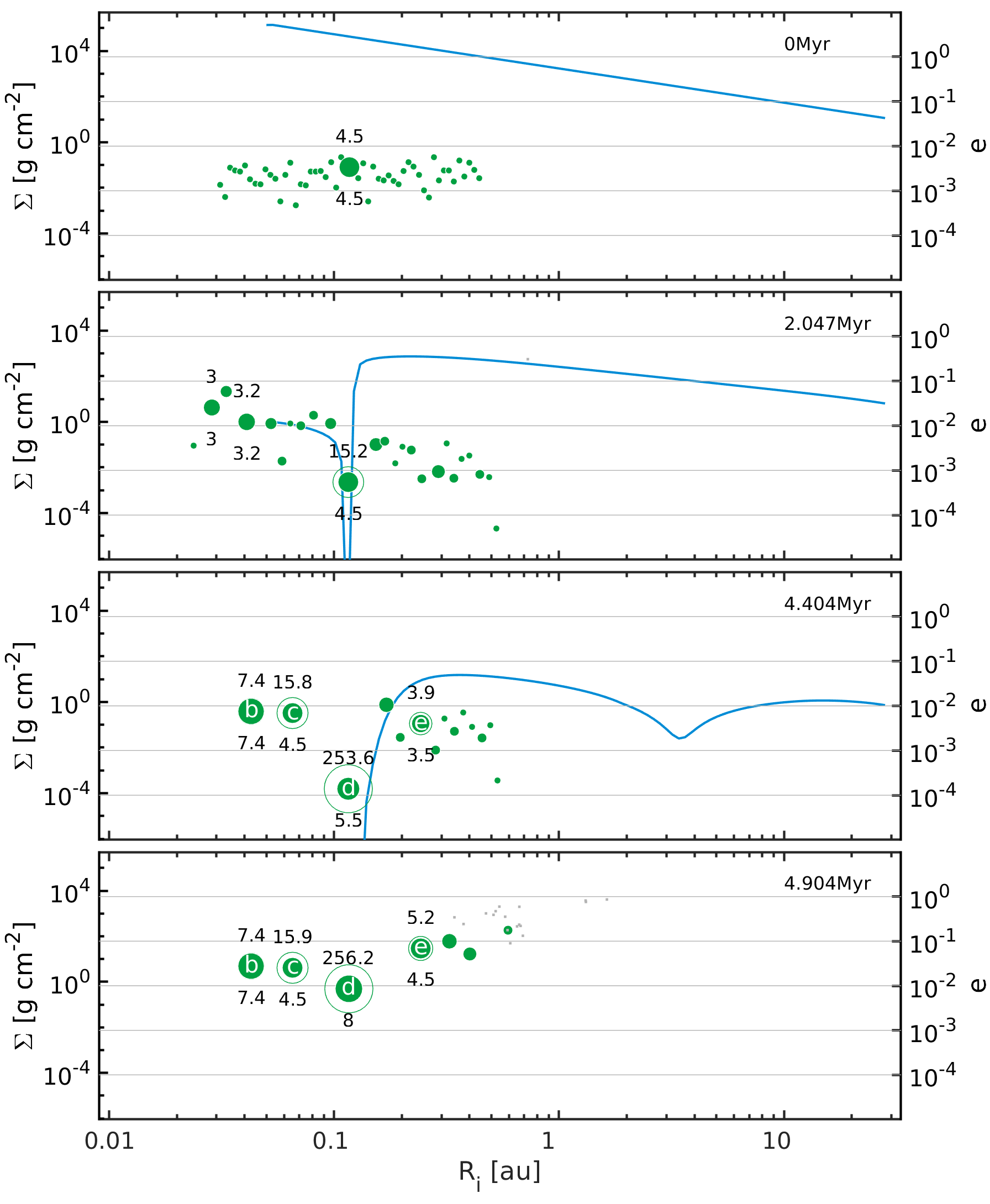}
 \caption[Giant formation in the seed-model scenario]{An example formation history from an instance of the the seed-model template \texttt{Kepler487-4.5M-5K} (same template as shown in figure \ref{fig:dym_evo_aeim}). The labels planet b, c, d, and e represent the same planets shown in figure~\ref{fig:dym_evo_aeim}. The blue lines are the disc surface density, green dots are the protoplanets, green circles are the gas envelope and grey dots are the collision debris. The sizes of the green dots and circles denote their relative mass. Protoplanet masses that are greater than $3\,\mathrm{M_{\oplus}}$ are shown by the text (upper: total mass; lower: core mass). Eccentricities are shown in the right vertical axes. The top panel shows the initial conditions, the second panel is the time that the first gap opens in the disc ($\sim 2.0\,\mathrm{Myr}$), the third panel is the time when photoevaporation starts to dominate the surface density evolution ($\sim 4.4\,\mathrm{Myr}$), and the bottom panel is the time after the disc has dissipated ($\sim 4.9\,\mathrm{Myr}$). 
 } \label{fig:1d-disc}
\end{center}
\end{figure}

\begin{figure}
\begin{center}
\includegraphics[width=\columnwidth]{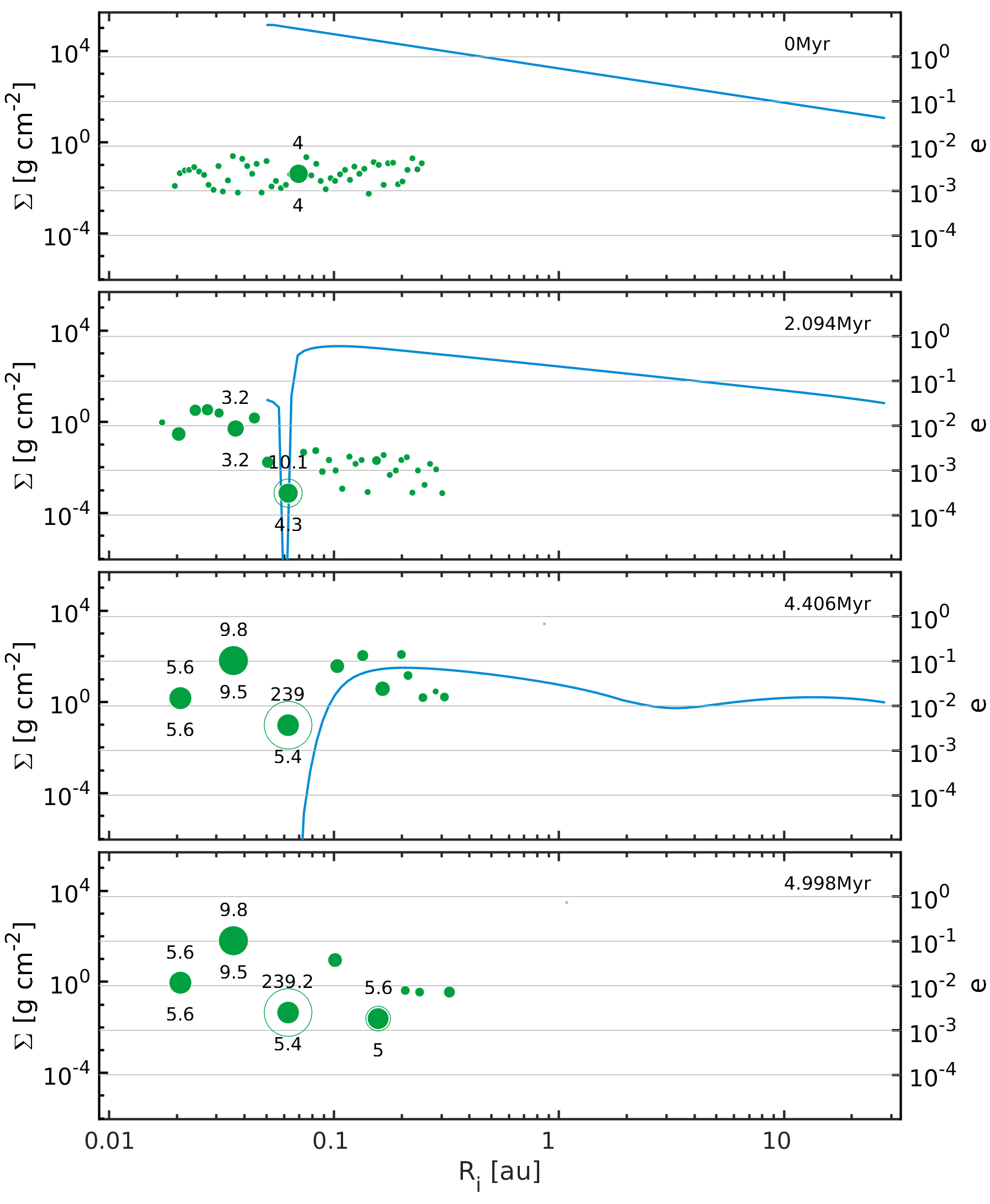}
 \caption[Disc and planet evolution for the seed-model \texttt{Kepelr730-4M-5k}]{Similar to figure~\ref{fig:1d-disc} but for another seed-model run from the \texttt{Kepler730-4M-5K} template. A closer-in giant planet is formed at $\sim 0.05 \,\mathrm{au}$, similar to the initial semimajor axis of the seed-protoplanet.}
 \label{fig:1d-disc_730}
\end{center}
\end{figure}

Figure \ref{fig:1d-disc} shows some additional instances from the run shown in figure \ref{fig:dym_evo_aeim} (i.e. from the \texttt{Kepler487-4.5M-5K} template), together with the evolution of the disc. The upper panel shows the initial conditions, where the blue line is the initial surface density (equation \ref{eq:initsigma}) and the green dots are the protoplanets. The second panel shows a moment at the beginning of the runaway gas accretion phase, where the seed has accreted enough mass to open a gap in the disc. The third panel shows the moment during the outer disc damping phase. By this time the inner super-Earths/sub-Neptunes were already formed while the outer protoplanets are still in a compact configuration due to the disc damping. The bottom panel shows the moment after the gas disc dissipates. Due to the lack of the disc damping forces, giant impacts between the outer embryos are common and allow the sub-Earth mass protoplanets to grow to super-Earths. Figure \ref{fig:1d-disc_730} shows a similar result to figure \ref{fig:1d-disc}, but for a run from the \texttt{Kepler730-4M-5K} template. The \texttt{Kepler730} templates follow similar behaviour to the \texttt{Kepler487} templates, and provide similar final architectures of the planetary systems. The giants formed in the \texttt{Kepler730} runs are closer-in at $\sim0.05$ to $0.06\,\mathrm{au}$ similar to the WASP-47, Kepler-730, and TOI-1130 systems, where all the giants formed have orbital periods less than 10 days.

\subsubsection{Multiplicities, masses, and orbital parameters}
There is a systematic difference, in terms of multiplicities and mass distributions, between the final inner and outer parts of the systems from the seed-model (see table \ref{tab:mul_mass}). The inner multiplicity, $N_{\mathrm{in}}$, always has a lower value than the outer multiplicity, $N_{\mathrm{out}}$. The inner total mass, $M_{\mathrm{total,in}}$, is similar to the outer total mass, $M_{\mathrm{total,out}}$, in the \texttt{Kepler730} and \texttt{TOI1130} templates, while the \texttt{Kepler487} and \texttt{Kepler89} templates hold a relation of $M_{\mathrm{total,in}}>M_{\mathrm{total,out}}$. This difference is because of the initial conditions of the \texttt{Kepler730} and \texttt{TOI1130} templates located most of the inner protoplanets between the host star and the inner edge of the disc, while the \texttt{Kepler487} and \texttt{Kepler89} templates are further out and allow most of the protoplanets to sit inside the disc when $K=5$. The difference between the two sets of $M_{\mathrm{total,in}}$, to a certain extent, shows how much gas mass the inner systems can accrete from the disc. A rough relation of $M_{\mathrm{total,in}}\approx 2M_{\mathrm{total,out}}$ can be drawn, which denoted a $\sim1:1$ solid-to-gas ratio for the inner systems.
\begin{table*}
\caption[Multiplicities and masses for the seed-model runs]{Final planetary multiplicities and masses for the seed-model runs. The subscripts `in' and `out' represent the inner part and the outer part of the system with respect to the location of the giant (the seed). $\left \langle M_{\mathrm{g}} \right \rangle$ is the average mass of the final giants, where the \texttt{4M} runs always give lower values than the \texttt{4.5M} runs, comparable to the final mass obtained in the single-planet case (figure \ref{fig:mass-init-fin}). \texttt{WASP47} runs are not included here because no giant was formed at the end of the simulations, for the reason described in section \ref{subsec:seedevo}. The table is separated according to the orbital radius of the giant planet, with the 4 upper templates having giants with $a<0.1\,\mathrm{au}$ and the 4 low templates having $a > 0.1\,\mathrm{au}$.}\label{tab:mul_mass}
\begin{tabularx}{0.9\textwidth}{L{1.35}R{0.9}R{0.9}R{0.95}R{0.95}R{1}R{1}R{0.95}}
\toprule\toprule 
Template       & $\left \langle N_{\mathrm{in}} \right \rangle$ & $\left \langle N_{\mathrm{out}} \right \rangle$ & $\left \langle M_{\mathrm{total,in}} \right \rangle$ & $\left \langle M_{\mathrm{total,out}} \right \rangle$ & $\dfrac{\left \langle M_{\mathrm{total,in}} \right \rangle}{\left \langle N_{\mathrm{in}} \right \rangle}$ & $\dfrac{\left \langle M_{\mathrm{total,out}} \right \rangle}{\left \langle N_{\mathrm{out}} \right \rangle}$ & $\left \langle M_{\mathrm{g}} \right \rangle$ \\
\midrule  
\texttt{Kepler730-4M}   & 1.9 & 3.9 & 13.4 & 20.6 & 7.1  & 5.3 & 297.9 \\
\texttt{Kepler730-4.5M} & 1.7 & 3.2 & 16.8 & 16.2 & 9.9  & 5.1 & 360.2 \\
\texttt{TOI1130-4M}     & 2.4 & 3.4 & 15.8 & 13.7 & 6.6  & 4.0 & 221.7 \\
\texttt{TOI1130-4.5M}   & 2.4 & 4.3 & 13.8 & 15.7 & 5.8  & 3.7 & 360.9 \\ 
\midrule
\texttt{Kepler487-4M}   & 2.4 & 2.9 & 34.0 & 16.5 & 14.2 & 5.7 & 178.6 \\
\texttt{Kepler487-4.5M} & 2.0 & 3.6 & 44.6 & 17.0 & 22.3 & 4.7 & 258.3 \\
\texttt{Kepler89-4M}    & 2.2 & 2.8 & 45.3 & 14.9 & 20.6 & 5.3 & 174.0 \\
\texttt{Kepler89-4.5M}  & 2.5 & 3.9 & 47.4 & 17.7 & 19.0 & 4.5 & 247.5 \\[5pt]
Overall                 & 2.2 & 3.5 & 28.9 & 16.5 & 13.2 & 4.8 & 262.4 \\
\bottomrule\bottomrule
\end{tabularx}
\end{table*}

The inner systems also show a higher average mass of the planets, $\left \langle M_{\mathrm{total,in}} \right \rangle/\left \langle N_{\mathrm{in}} \right \rangle$, compared to the average planet mass of the outer systems, $\left \langle M_{\mathrm{total,out}} \right \rangle/\left \langle N_{\mathrm{out}} \right \rangle$. The inner average masses of the \texttt{Kepler487} and \texttt{Kepler89} templates are comparable to Neptune's mass, and the template with closer-in seeds, \texttt{Kepler730} and \texttt{TOI1130}, are more likely to host super-Earths.

\begin{figure}
\begin{center}
\includegraphics[width=\columnwidth]{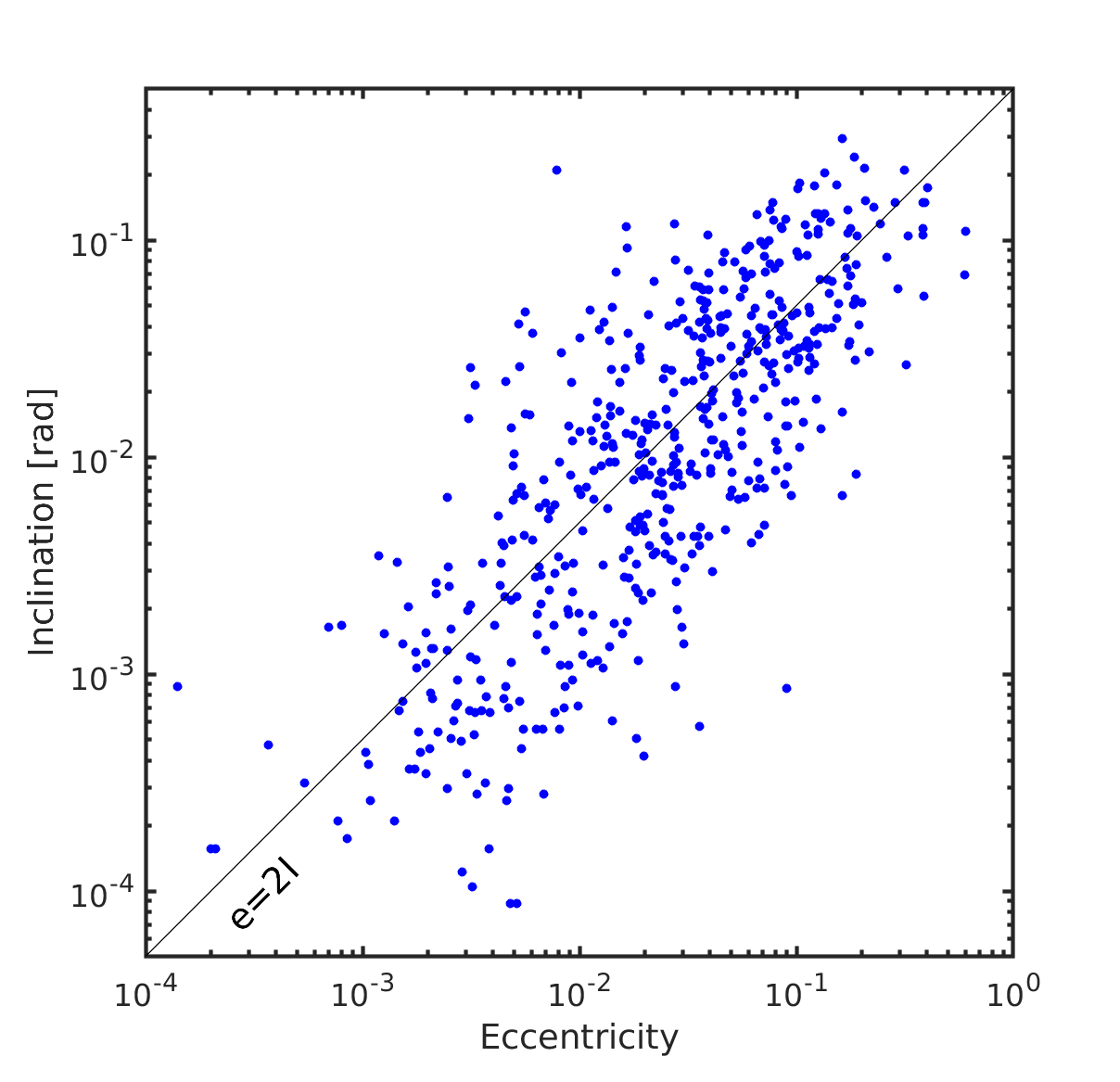}
 \caption[Final $e$-$I$ relation]{The relation between the final eccentricities and inclinations from the seed-model. The black line represent the relation of $e=2I$. The eccentricity and inclination distributions provide a good fit to the $e=2I$ relation, showing a signal of dynamical relaxation.}
 \label{fig:e-i-relation}
\end{center}
\end{figure}

There are no systematic differences in the final orbital parameters when comparing the \texttt{4M-5K} and \texttt{4.5M-5K} runs, but a clear divergence emerges between the giants and other planets. As shown in figure~\ref{fig:dym_evo_aeim}, \ref{fig:1d-disc} and \ref{fig:1d-disc_730}, the giants retain low eccentricities throughout the simulations. The final mean eccentricity of the giants has a value of $\left \langle e \right \rangle \approx 0.01$, while the other planets yield $\left \langle e \right \rangle \approx 0.06$. And the final mean inclination for the giants is $\left \langle I \right \rangle \approx 0.005\,\mathrm{rad}$ and $\left \langle I \right \rangle \approx 0.03\,\mathrm{rad}$ for the companions. The final outcomes of the simulations shows a strong correlation between the distribution of the eccentricities and inclinations (figure \ref{fig:e-i-relation}). This shows that the systems have undergone dynamical relaxation \citep{2005dpps.conf...41K}, as expected.

\subsection{Comparison to the equal-mass models}\label{subsec:comp_emm}
Two sets of simulations were performed for the equal-mass model, one with mutual separations $K=4$ (\texttt{XM-4K}), and one with $K=5$ (\texttt{XM-5K}). 
Without a seed-protoplanet in the simulations, the formation of a giant planet is expected to occur less frequently than in the seed-model. Forming a gas giant requires a sufficiently massive planet to form early in the gas disc lifetime so it has time to undergo runaway gas accretion before the disc disperses, and this clearly requires a planet to undergo numerous mergers.

\begin{figure}
\begin{center}
\includegraphics[width=\columnwidth]{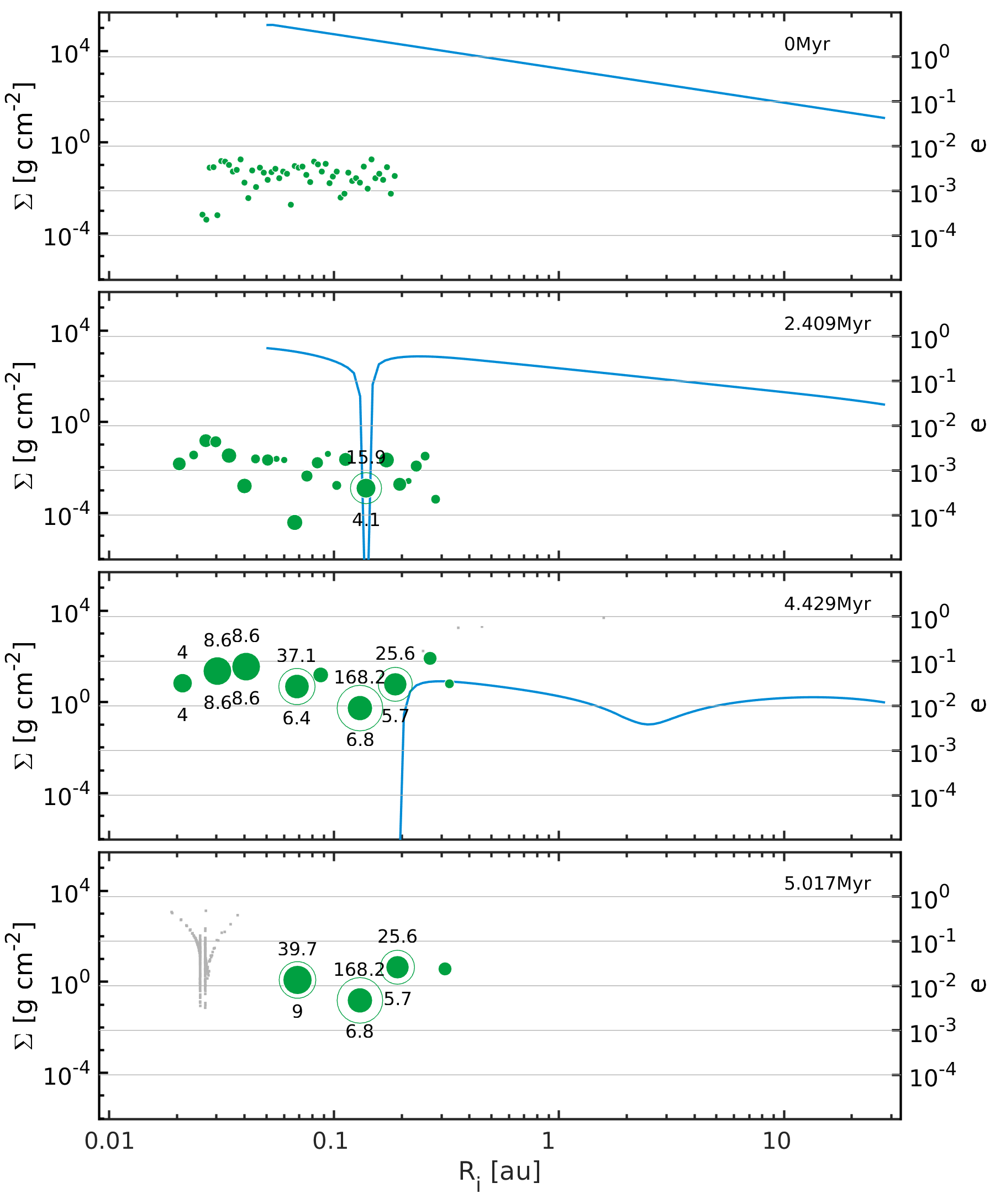}
 \caption[Disc and planetary evolution of the equal-mass-model - \texttt{Kepelr730-XM-4k}]{Similar to figure \ref{fig:1d-disc} but for a run of the equal-mass \texttt{Kepler730-XM-4k} template. The gap opening, photoevaporation, and disc dissipation times are similar to the seed-model runs. Unlike the seed-model run (figure \ref{fig:1d-disc_730}), the formation of giant occurs at a different location to the reference location of the \texttt{Kepler730} giant planet.}
 \label{fig:1d-disc-xmg}
\end{center}
\end{figure}

Figure~\ref{fig:1d-disc-xmg} shows an example of a run that forms a giant planet. A large core with $4.1\,\mathrm{M_{\oplus}}$ forms and enters the runaway gas accretion stage at $\sim2.4\,\mathrm{Myr}$, so this protoplanet has enough time to undergo runaway growth and transition to accreting gas at the viscous supply rate for $\sim 2\,\mathrm{Myr}$, eventually becoming a gas giant. Another feature we can see in the bottom panel of figure~\ref{fig:1d-disc-xmg} is the large amount of debris at the inner edge of the system. The debris were formed during a super-catastrophic collision event. It was caused by the dynamical instability and giant impact between two protoplanets during and shortly after the disc dissipation (the late chaotic phase).

\begin{figure}
\begin{center}
\includegraphics[width=\columnwidth]{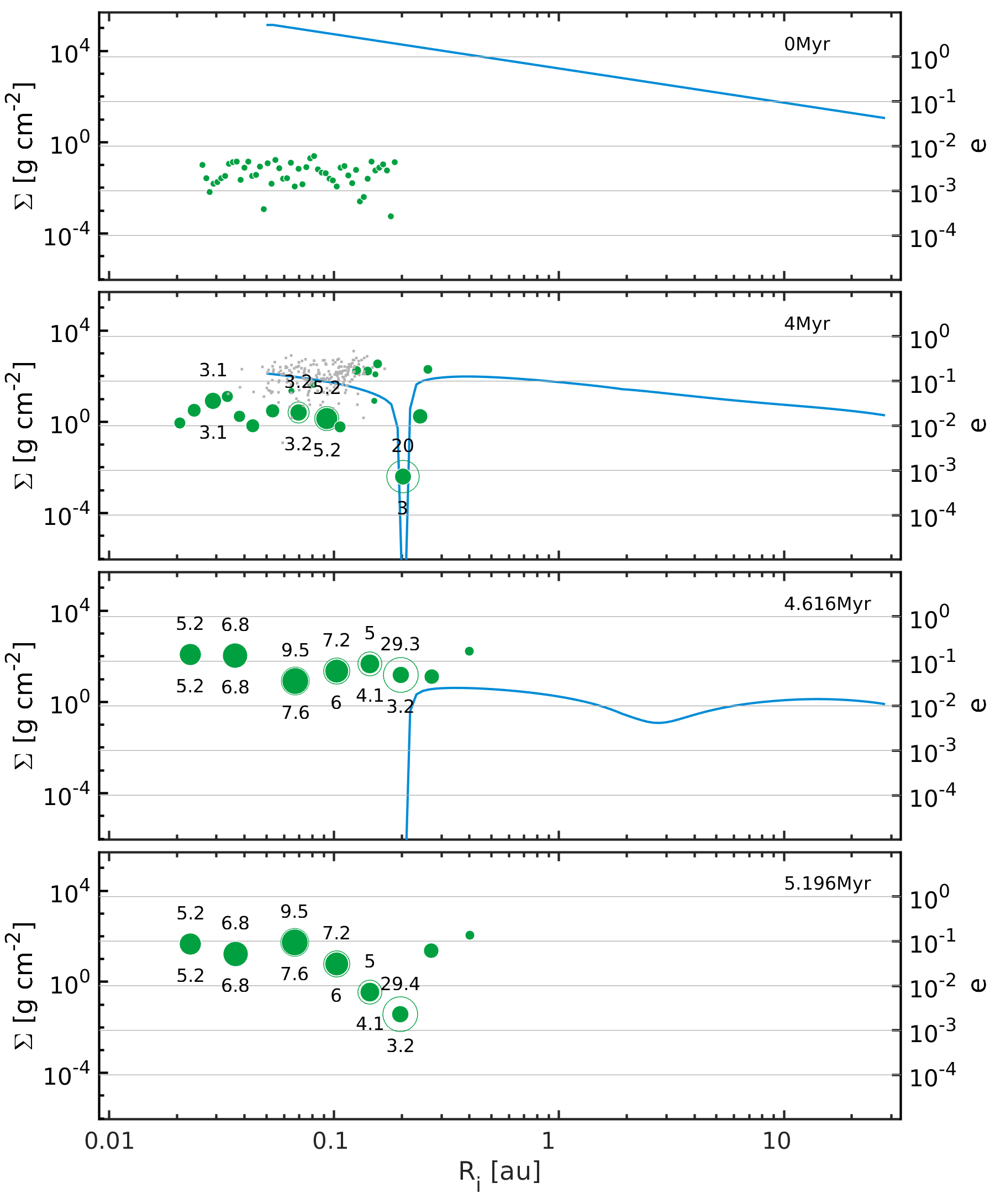}
 \caption[Dynamical history of the equal-mass-model (no giant) - \texttt{Kepelr730-XM-4K}]{Similar to figure \ref{fig:1d-disc} and \ref{fig:1d-disc-xmg} but for another run of the \texttt{Kepler730-XM-4K} template. This equal-mass run does not result in the formation of a giant, which is a common outcome in the equal-mass models. The transition to runaway gas accretion and gap opening occurs at $\sim4\,\mathrm{Myr}$, leaving insufficient time for the protoplanet to accrete a massive envelope.}
 \label{fig:1d-disc-xm-nog}
\end{center}
\end{figure}

Figure \ref{fig:1d-disc-xm-nog} demonstrates a case where no gas giant forms. The largest protoplanet in this run only enters the runaway gas accretion phase and opens a gap at $\sim4\,\mathrm{Myr}$, which is too close to the end of the disc lifetime for it to form a Jovian mass planet.

The formation of gas giants is not common in our equal-mass models. Only 1\% of our equal-mass runs (all of them from the $K=4$ runs) produce a gas giant with final mass greater than $100\,\mathrm{M_{\oplus}}$. We note that the giant formation percentage is sensitive to the gas accretion rate (equation \ref{eq:accretionrate}), and we will discuss the effect of considering different gas accretion prescriptions in section~\ref{sec:opa}.

The dynamical evolution observed in the equal-mass models are, in general, different from the evolution history of the seed-model (section \ref{subsec:seedevo}), except for the runs in which a giant was formed. For those few runs with giant formation, they follow the dynamical evolution path of the seed-model. More commonly, the equal-mass runs evolve in three phases: \emph{early giant impacts}, \emph{disc damping}, and the \emph{late chaotic phase}. The first and final phases are the same as the seed-model. The disc damping phase is the extension of the early impact phase, where the damping of the eccentricities and inclinations is the dominant effect on the orbital evolution of the protoplanets. 

\begin{figure}
\begin{center}
\includegraphics[width=\columnwidth]{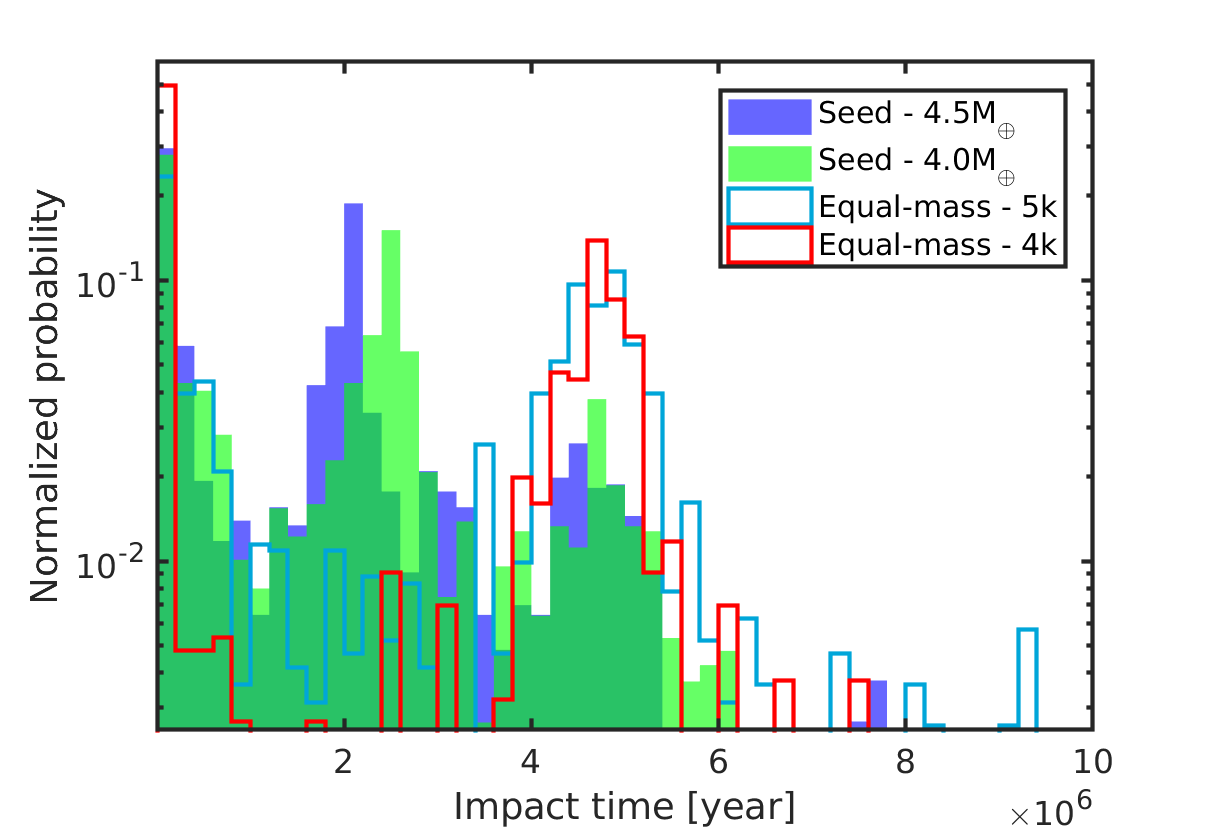}
 \caption[Time of giant impacts]{Normalized probabilities of all the giant impact events with respect to time during the four different set of simulations, including the \texttt{4.5M} seed-model (blue area), \texttt{4M} seed-model (green area), \texttt{5k} equal-mass-model (cyan line), and \texttt{4k} equal-mass-model (red line).}
 \label{fig:impacttime}
\end{center}
\end{figure}

Figure~\ref{fig:impacttime} shows the comparison of the impact time between the four subsets of our seed-model and equal-mass model. It is clear that all four of our subsets have an early impact phase. The $K=5$ systems follow a similar rate of decrease at the number of impacts ($<1\,\mathrm{Myr}$). The $K=4$ systems (red line) show the majority of early impacts happened before $0.2\,\mathrm{Myr}$, as an effect of a more compact configuration initially. In the \texttt{XM-4K} runs, the equal-mass protoplanets can merge with each other to grow sufficiently massive to accrete a noticeable gas envelope earlier than the \texttt{XM-5K} runs. This explains why the \texttt{XM-4K} runs, compared to the \texttt{XM-5K} runs, are more likely to form a giant planet.

The seed-model runs (\texttt{4M-5K} and \texttt{4.5M-5K}) have peaks between $\sim 2.0$ to $2.4\,\mathrm{Myr}$, corresponding to the runaway gas accretion phase. This peak comes earlier for the \texttt{4.5M-5K} set (blue filled bars) at $2.0\,\mathrm{Myr}$ than the \texttt{4M-5K} set (green filled bars) at $2.4\,\mathrm{Myr}$. This is simply because the higher mass seeds undergo runaway gas accretion earlier than the lower mass seeds. 

All four sets have a peak in the collision times at $\sim 5\,\mathrm{Myr}$, which is the average lifetime of our disc models. This marks the onset of the late chaotic phase, and arises because the eccentricity and inclination damping forces diminish as the disc disperses. The peaks for the equal-mass sets are higher than the seed sets. This results from the extended disc damping phase for the equal-mass models, where giant impacts are much more common after the disc disperses instead of during the lifetime of the disc. During the gas-free stage, the collisions are more frequent within a million years after the disc has dissipated, and the systems became more dynamically quiet at $\sim 6\,\mathrm{Myr}$ and beyond.

\subsection{Collision behaviour}
\begin{figure}
\begin{center}
\includegraphics[width=\columnwidth]{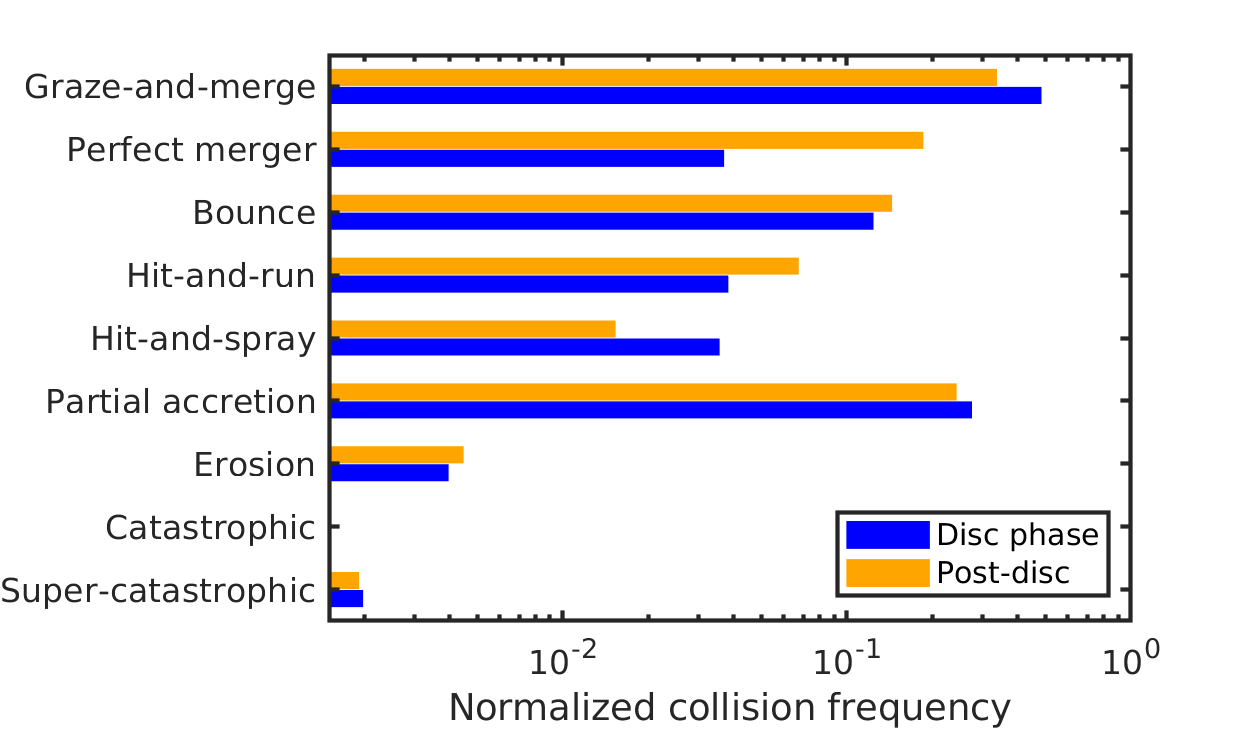}
 \caption[Collision types]{The frequency of each collision type during the two different phases of the simulations. The collisions happening during the disc phase are recorded in the blue bars. The collisions occurring after disc dispersal are recorded in orange bars.}
 \label{fig:collisiontype}
\end{center}
\end{figure}

The outcomes of protoplanet collisions show some differences when they occur during gas disc phase and after the disc dispersed. Figure~\ref{fig:collisiontype} reveals the frequency of different types of collisions. The blue bars indicate the collision frequency during the disc phase, and the orange bars shows the frequency after the disc is no longer present. 

In general, the type of collision that can create a large amount of debris, such as super-catastrophic, catastrophic, and erosion collisions, are not common. For the post-disc phase, the most common type of collision is graze-and-merge, followed by partial accretion. The collision frequency results for the post-disc phase is in strong agreement with the gas-free simulations by \citet{2020MNRAS.491.5595P} and \citet{2020MNRAS.493.4910S}, where these two studies also considered the same collision model.

\begin{figure}
\begin{center}
\includegraphics[width=\columnwidth]{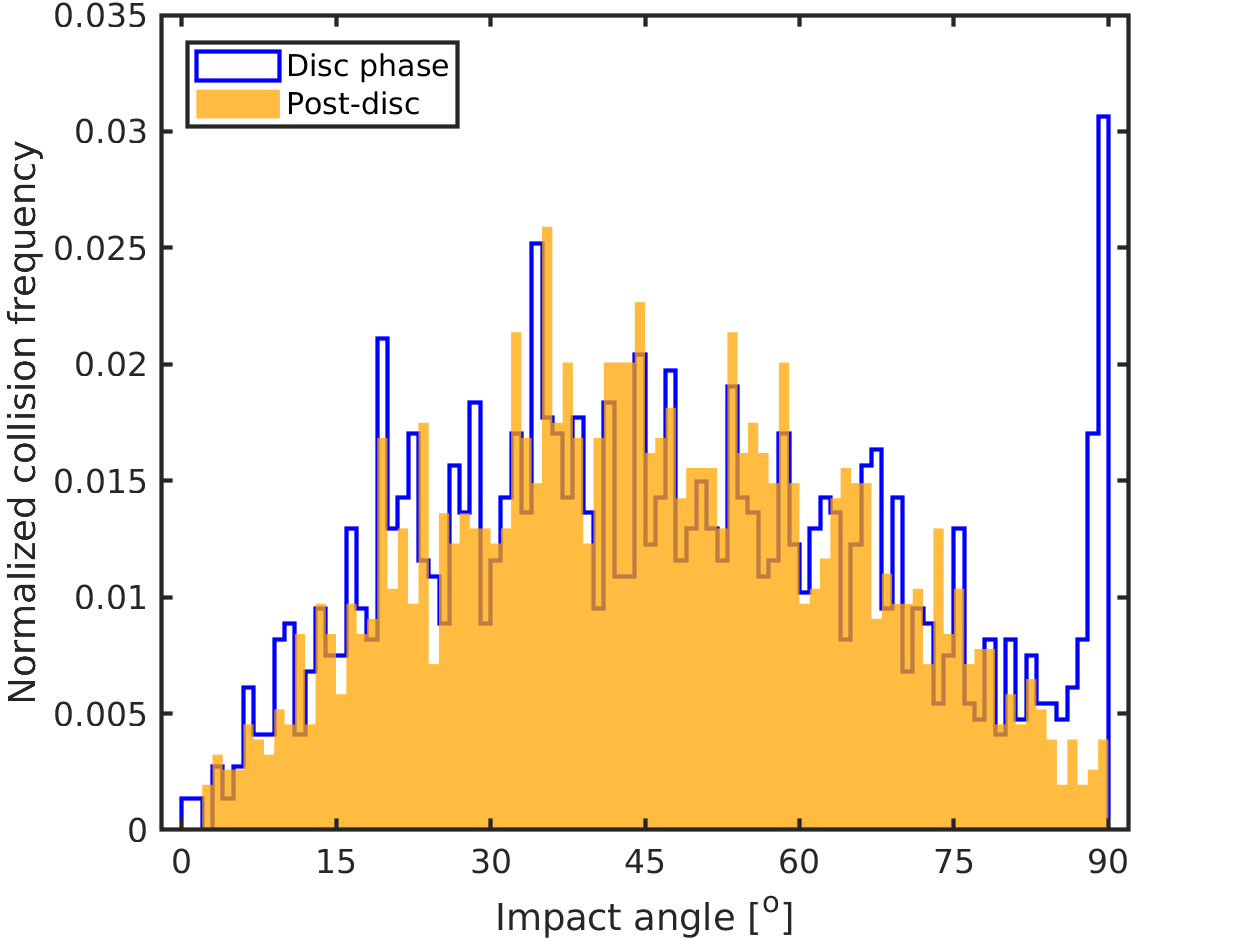}
 \caption[Collision angle]{The frequency of impact angles of all the giant impact events in our simulations. The collisions happening during the disc phase are plotted in blue. The collisions occurring after disc dispersal are plotted in orange. The peak of the disc phase collisions occurring close to $90^{\circ}$ is the result of collisions between pairs of planets that become gravitationally bound to each other, and which then spiral in before colliding, leading to graze-and-merge collisions.}
 \label{fig:collisionangle}
\end{center}
\end{figure}

The collision behaviour during the disc phase is different, and almost half of the collisions are graze-and-merge. Together with the drop of the number of perfect mergers, it indicates that there are more slow collisions with high impact angles compared to the gas-free stage. This phenomenon can also be noticed in figure \ref{fig:collisionangle}, where the figure illustrates the frequencies of different impact angles. The distribution of impact angles for the post-disc stage (orange filled bars) follow a distribution peaking at around $45^{\circ}$ \citep{2020MNRAS.491.5595P,2020MNRAS.493.4910S}. Meanwhile, the impact angles in the disc phase (blue line) follow a similar distribution, except in the large angle domain. There is a sharp increase of the frequency at impact angles between $85^{\circ}$ to $90^{\circ}$. This indicates that the disc influences the dynamics and induces slow mergers between protoplants at high impact angle. The high impact angles show that the orbits of the two colliding bodies are more circular, instead of arising from high eccentric orbital crossing. Analysis of the results shows that these collisions actually arise because pairs of planets become gravitationally bound to each other when the disc damping is present, and these planets spiral in before colliding with high impact angles.

\subsection{Observational detection rate hot Jupiter and inner super-Earth systems}\label{subsec:synthetic}
Transit surveys have discovered $\sim2500$ planetary systems. Among this population, five systems contain a transiting giant with orbital period less than 30 days and transiting super-Earths/mini-Neptunes (section~\ref{sec:setup}) that orbit interior to the giant. These systems are WASP-47, Kepler-730, TOI-1130, Kepler-487, and Kepler-89. These systems make up $\sim0.2\%$ of the whole population of transiting systems. To compare our simulation outcomes with this detection rate, we carried out synthetic transit observations of our final planetary systems.

Each simulated planetary system is synthetically observed from 100,000 randomly chosen viewing locations, isotropically distributed with respect to each host star. To compare our results to transit surveys, we only consider planets that satisfy the observation limits of a Kepler-like survey. Therefore, we only consider planets, and exclude all collision debris, with orbital radius less than $1\,\mathrm{au}$. 

All our seed-model runs (except \texttt{WASP47}) contain a hot Jupiter at the end of the simulations. Synthetic transit observations of the final seed-model system tell us that $30.2\%$ of planetary systems that are detected contain a hot Jupiter and an inner super-Earth. In this model, when a giant planet is detected, there is only a $\sim5\%$ chance that the inner companion will not also be picked up as a transiting planet.

\begin{figure}
\begin{center}
\includegraphics[width=\columnwidth]{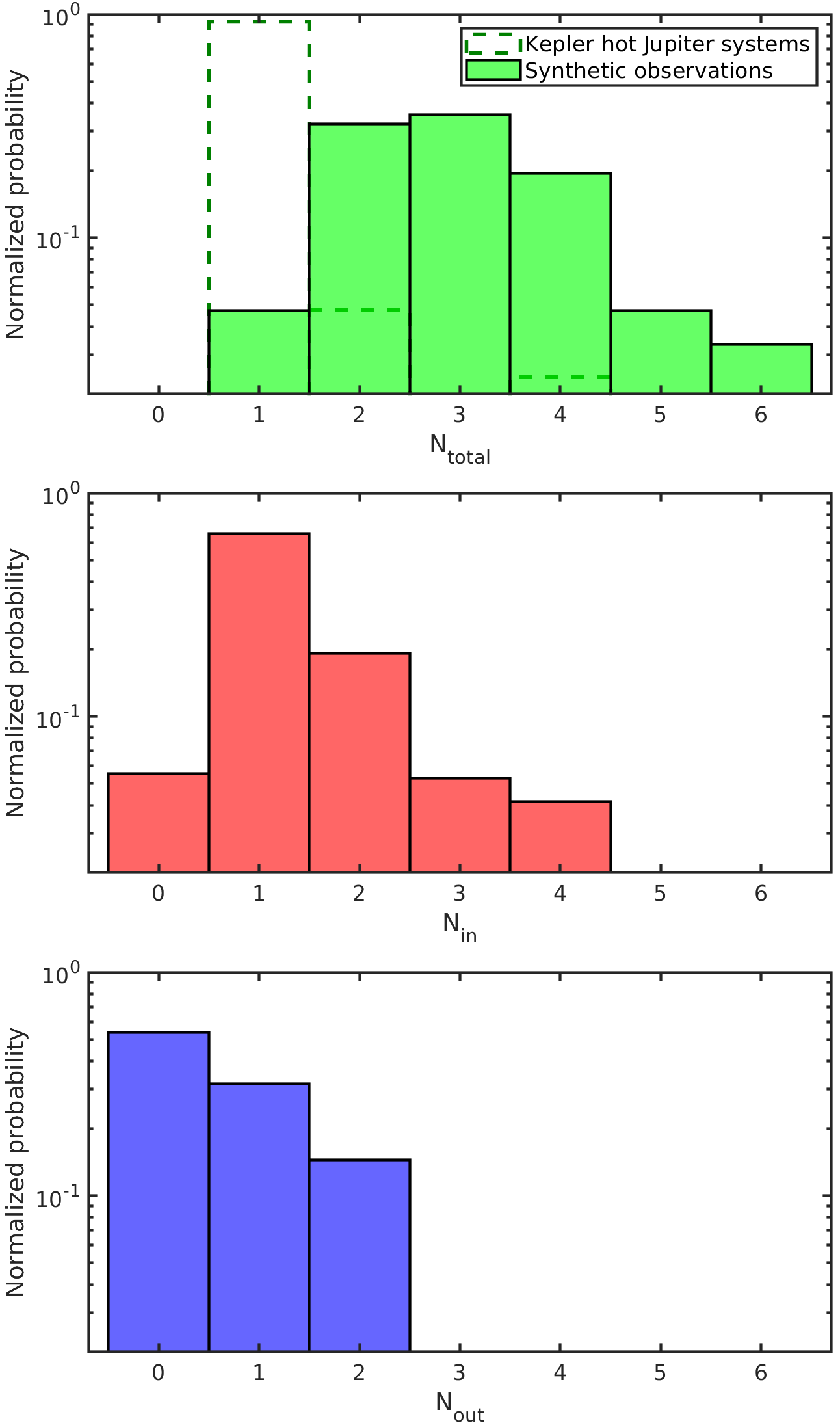}
 \caption[Synthetic observed multiplicities of hot-Jupiter systems]{Normalized probabilities of the final multiplicity, $N$, of the synthetically observed hot Jupiter systems arising from the equal-mass model. The subscript `total' represents the total multiplicity of the systems (top panel). The bars marked using dashed lines are the hot Jupiter systems observed by Kepler, and those filled with light green are the synthetically detected hot Jupiter systems from the simulations. The value of $N_{\mathrm{total}}=1$ corresponds to only the giant in the systems being detected. The subscripts `in' (middle panel) and `out' (bottom panel) represents the numbers of planet detected interior or exterior to the giant.}
 \label{fig:Ntotalinout}
\end{center}
\end{figure}

The equal-mass runs also show there is only a $\sim5\%$ chance that an inner companion will not be detected when a giant is detected in synthetic observations (figure \ref{fig:Ntotalinout}), similar to the seed-model. Hence, figure~\ref{fig:Ntotalinout} shows that a prediction of these \textit{in situ} formation simulations is that hot Jupiters detected in transit surveys should almost always be detected with interior super-Earths, and they should be detected with exterior planets about 50\% of the time, which is clearly not the case. Hence the initial conditions we have adopted, or some other aspects of the model, do not apply to the majority of hot Jupiter systems.

To recap, approximately $1\%$ of the equal-mass simulations produce a giant planet (section \ref{subsec:comp_emm}). Synthetic transit observations of all of our equal-mass simulations show that the proportion of detected planetary systems containing both a hot Jupiter and an inner super-Earth is $0.26\%$. This is similar to the occurrence rate of $\sim0.2\%$ from the actual transit observations. There are five systems (as we defined and selected) containing a hot Jupiter and an inner transiting planet within a total number of $\sim2500$ transiting exoplanet systems. Nonetheless, in spite of this area of agreement, it is clear the model presented here does not accord with the majority of hot Jupiter systems observed in transit surveys that are able to detect super-Earth companions, as shown in figure~\ref{fig:Ntotalinout}. The \textit{in situ} model produces too few systems in which a hot Jupiter would be detected as a single planet, unaccompanied by a super-Earth companion.

\section{Impact of varying gas envelope accretion prescription}\label{sec:opa} 
The gas envelope accretion routine is an important ingredient that affects the giant planet formation rate. In section~\ref{sec:mainresult}, we adopted equation \ref{eq:accretionrate} to accrete gas envelopes during the phase when the planet is embedded in the disc, and this prescriptions is insensitive to local disc conditions as it is based on calculations performed at 5.2\,au (see figure \ref{fig:mass-init-fin}, dashed lines). Realistically, the local gas accretion rate is dependent on local disc properties, such as the local temperature and opacity. In this section, we investigate the impact of adopting equation~\ref{eq:gasenvelope-gc}, which depends on local disc conditions, on the planetary systems formed in the simulations.

\subsection{Simulation outcomes}
To investigate the changes to our main results (section~\ref{sec:mainresult}) that arise when using a more realistic gas envelope accretion model, we ran an extra set of equal-mass simulations with equation \ref{eq:gasenvelope-gc}. Equation~\ref{eq:gasenvelope-gc} does not have a finite solution if $M_{\mathrm{ge}}=0$ due to the negative power index, hence we initialise the protoplanets with a small and dynamically negligible envelope with $M_{\mathrm{ge}}=10^{-8}\,\mathrm{M_{\oplus}}$ for all the protoplanets in the simulations. We consider three sets of runs with different values of the opacity reduction factor $f_{\mathrm{opa}}=10^{-1}$, $10^{-2}$, and $10^{-3}$, and we use the labelling convention \texttt{o1}, \texttt{o2}, and \texttt{o3} when describing the simulations. Together with the initial $K$-values ($K=4$ and $5$) and the five selected systems, there are a total of 30 templates (3 sets of $f_{\mathrm{opa}}$ $\times$ 2 sets of initial $K$-value $\times$ 5 selected systems).

Equation~\ref{eq:gasenvelope-gc} produces a larger gas accretion rate for a smaller value of $f_{\mathrm{opa}}$, as expected. We find that the \texttt{o1} and \texttt{o2} runs do not produce any giants with mass $>100\,\mathrm{M_{\oplus}}$, while the \texttt{o3} runs have a giant formation rate of $\sim 6 \%$. Hence, compared to the giant formation rate of the equal-mass runs in the main result (section~\ref{subsec:comp_emm}), the \texttt{o3} runs produce more giants while the \texttt{o2} runs produce fewer. Synthetically observing the \texttt{o3} runs as described in section~\ref{subsec:synthetic} yields a $1.36\%$ chance that a detected system contains a transiting hot Jupiter and at least one transiting inner super-Earth.

\begin{figure}
\begin{center}
\includegraphics[width=\columnwidth]{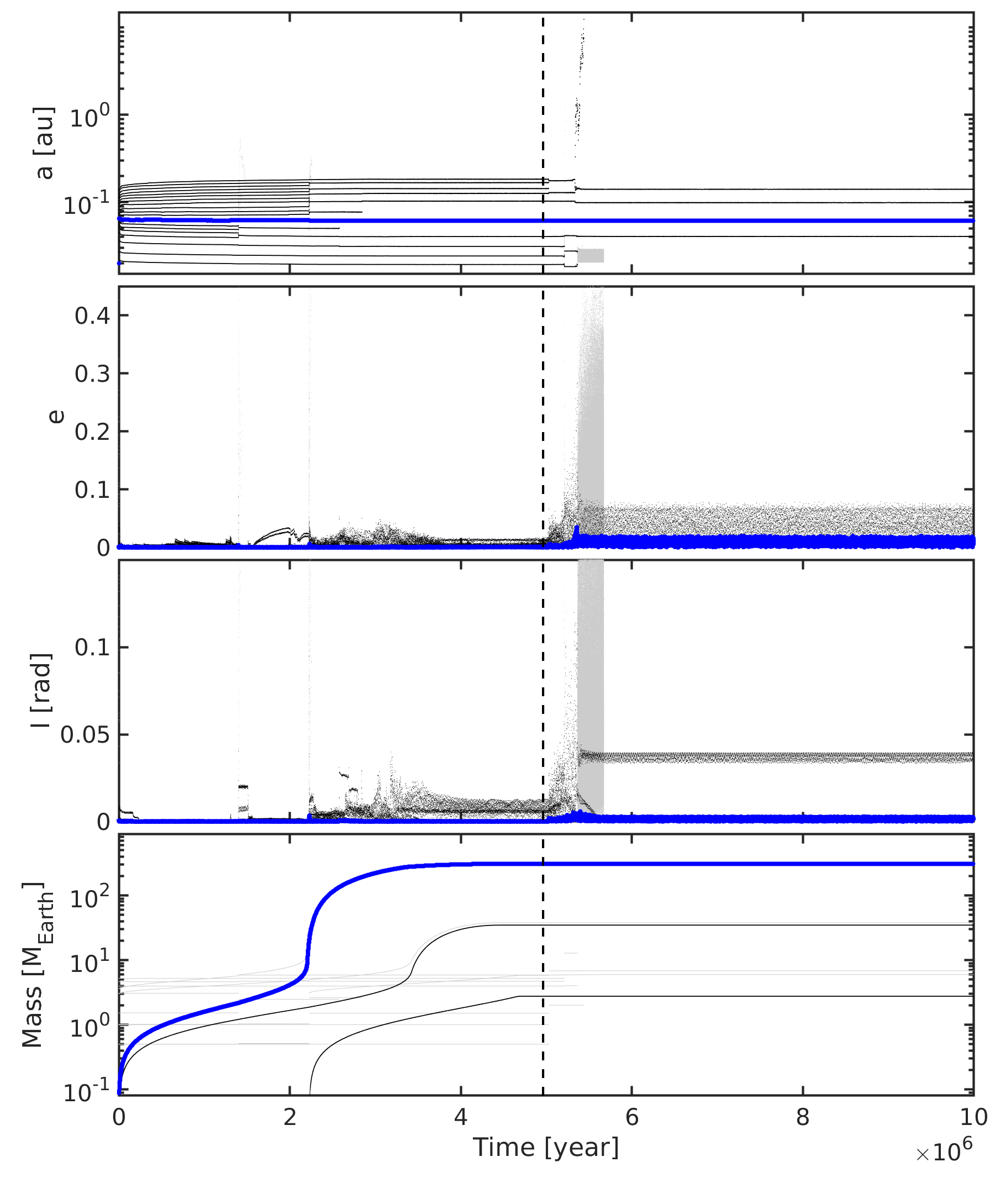}
 \caption[Dynamical evolution of the \texttt{WASP47-o3-4K} template]{Similar to figure \ref{fig:dym_evo_aeim} but for a run of the \texttt{WASP47-o3-4K} template, showing the evolution of the semimajor axes, eccentricities, inclinations, and masses. The blue lines indicated the most massive object in the simulation.}
 \label{fig:WASP47o34k_aeim}
\end{center}
\end{figure}

\begin{figure}
\begin{center}
\includegraphics[width=\columnwidth]{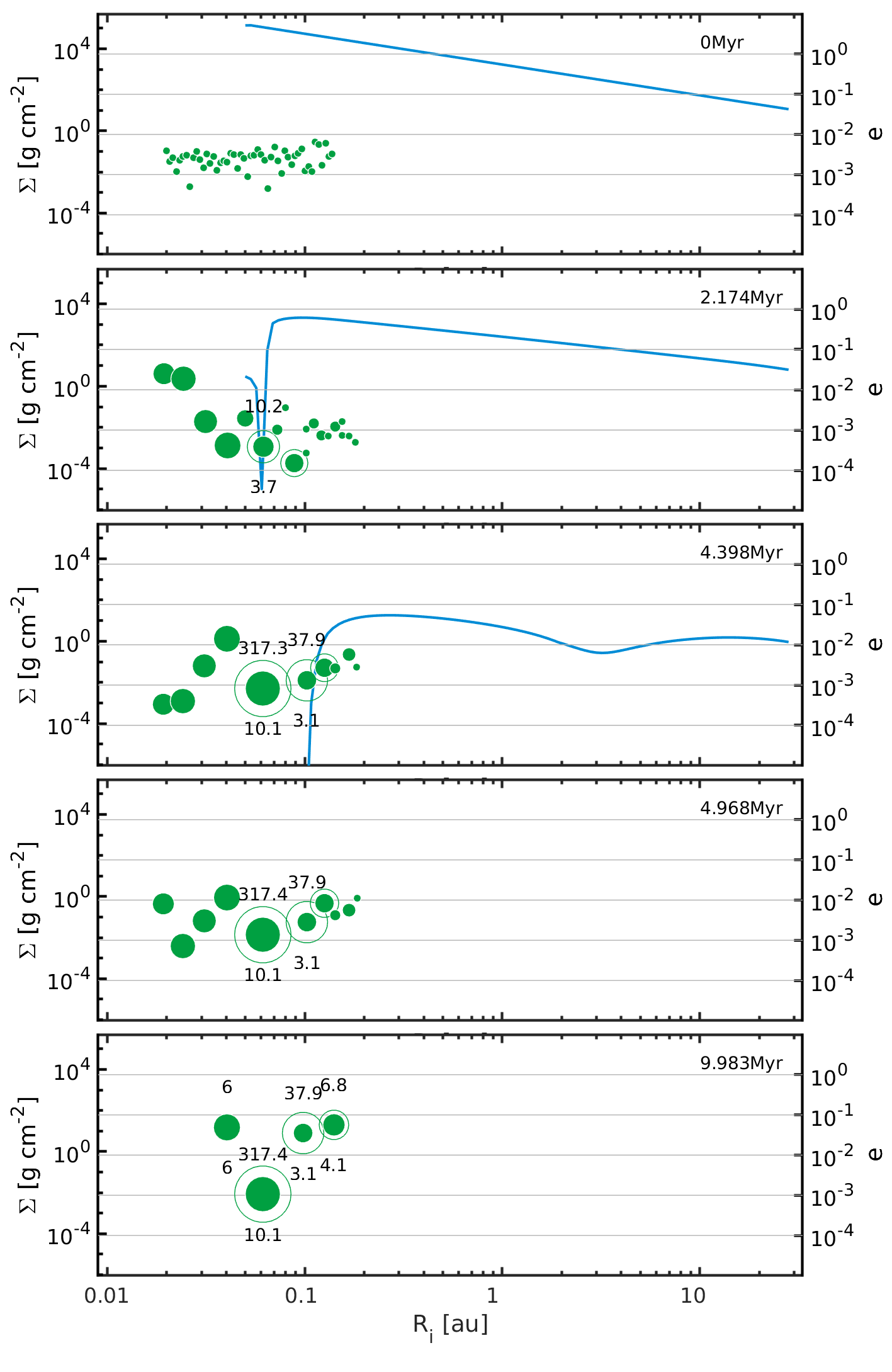}
 \caption[Disc and planets evolution of the \texttt{WASP47-o3-4k} template]{Similar to figure \ref{fig:1d-disc} but for the same run as shown in figure \ref{fig:WASP47o34k_aeim}, an equal-mass run with the new gas envelope accretion routine. As with the other equal-mass runs that form a giant planet, the gap opening, photoevaporation, and disc dissipation times are similar to the seed-model runs. The bottom panel shows a time close to the end of the simulation, and the system has a hot Jupiter at $\sim0.06\,\mathrm{au}$ and one inner super-Earth.}
 \label{fig:WASP47o34k_disc}
\end{center}
\end{figure}

Figures \ref{fig:WASP47o34k_aeim} and \ref{fig:WASP47o34k_disc} show the dynamical evolution of a run from the \texttt{WASP47-XM-4K-o3} template, where a hot Jupiter orbiting at $\sim0.06\,\mathrm{au}$ is formed with an inner super-Earth orbiting at $\sim0.04\,\mathrm{au}$. The \texttt{o3} runs which successfully formed one hot Jupiter, follow a very similar evolution as our main set of simulations (e.g. figure \ref{fig:1d-disc-xmg}) and always contain an inner planet.

\begin{figure}
\begin{center}
\includegraphics[width=\columnwidth]{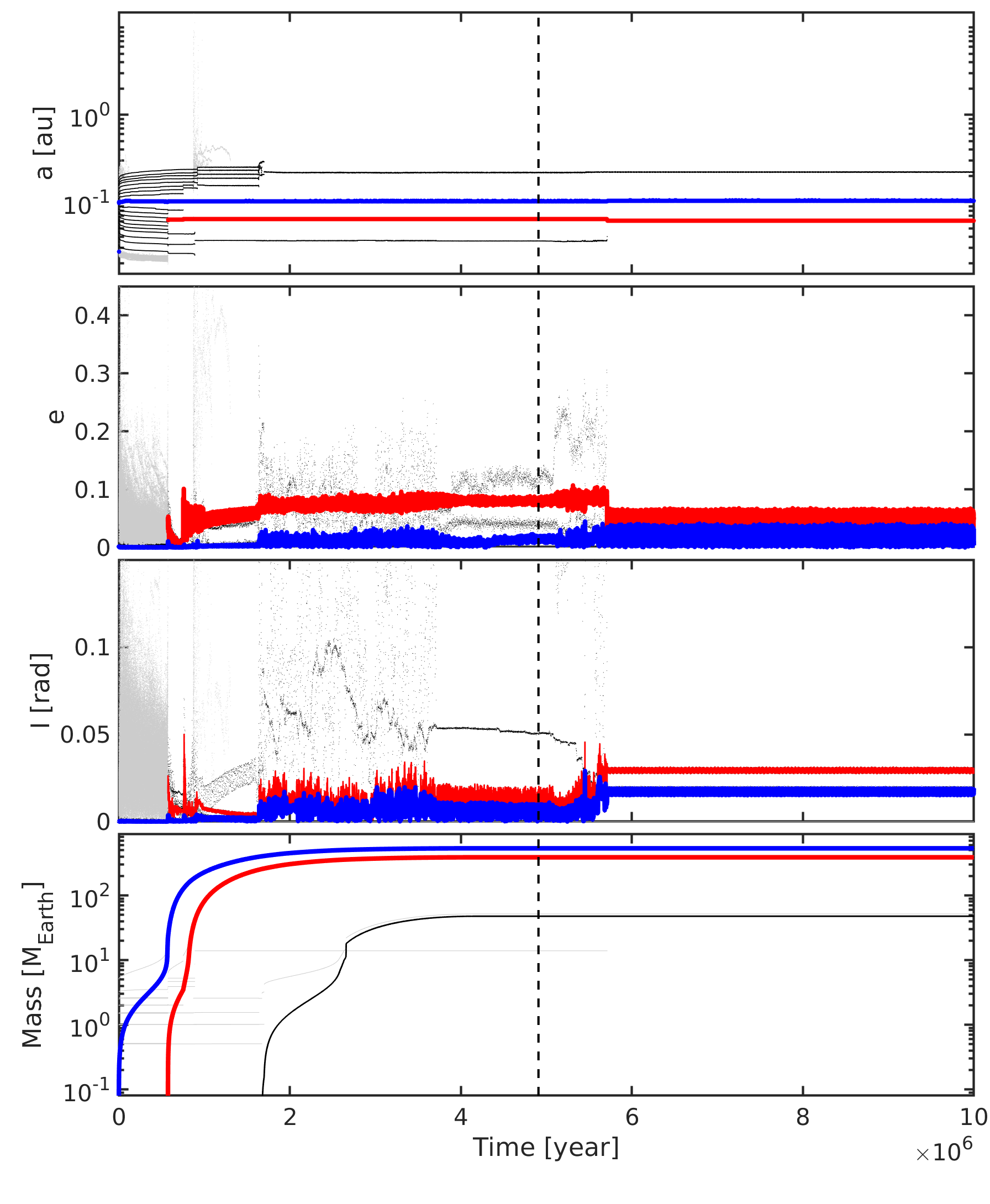}
 \caption[Dynamical history of \texttt{Kepler730-o3-4k} - 2 giants formation]{Evolution of a run from the \texttt{Kepler730-o3-4K} template, similar to figure~\ref{fig:dym_evo_aeim}. The blue lines indicate the most massive planet the simulation, and the red lines indicated the second-most massive object.}
 \label{fig:2g_aeim}
\end{center}
\end{figure}

\begin{figure}
\begin{center}
\includegraphics[width=\columnwidth]{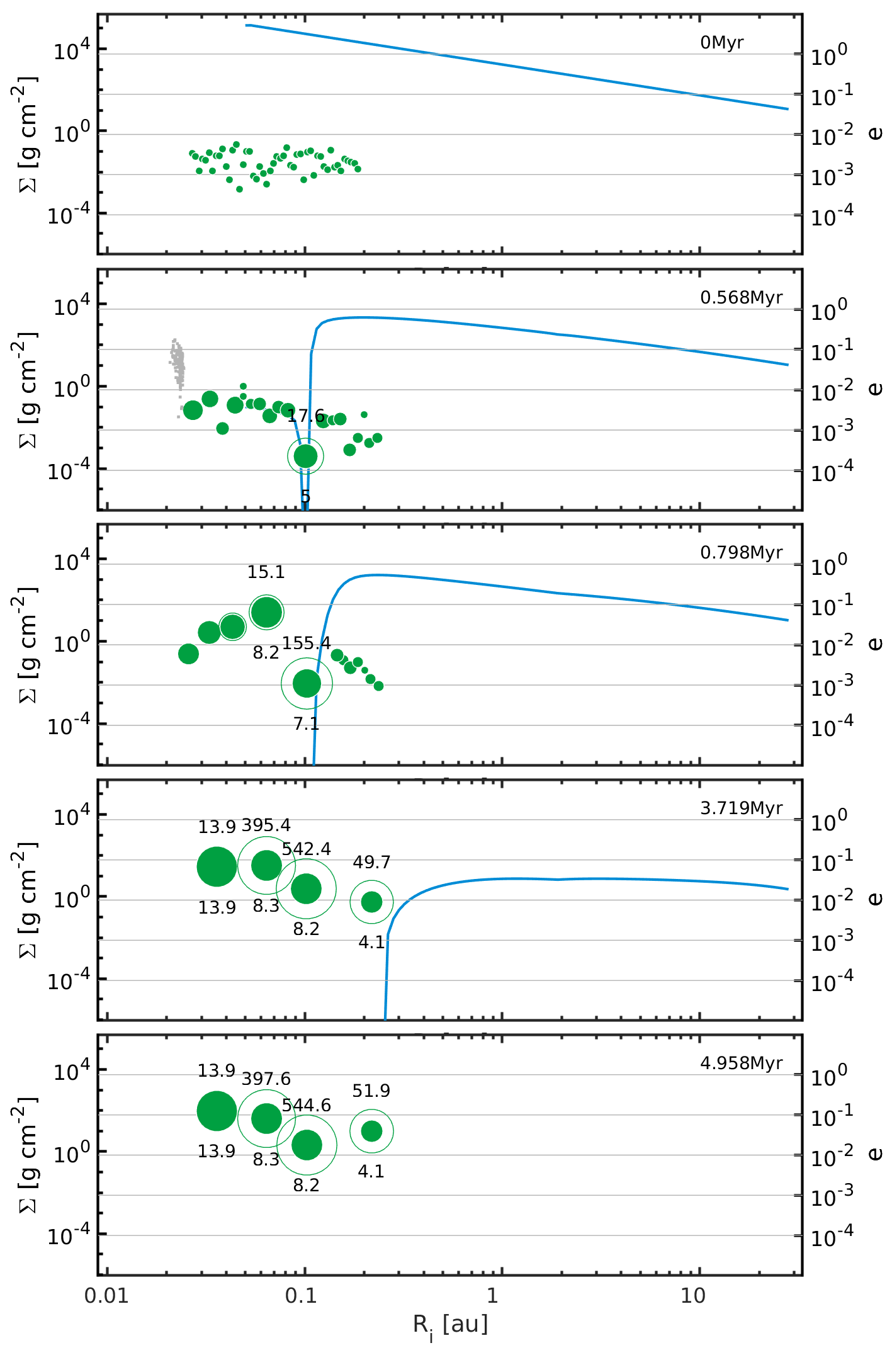}
 \caption[Disc and planets evolution of the \texttt{Kepler730-o3-4k} template]{Similar to figure~\ref{fig:WASP47o34k_disc} but for the same run as shown in figure \ref{fig:2g_aeim}. The first gap opening planet forms early ($\sim 0.5\,\mathrm{Myr}$). Another large core forms before $1\,\mathrm{Myr}$, allowing two gas giants to form before the gas disc dissipates.}
 \label{fig:2g_disc}
\end{center}
\end{figure}

Interestingly, $\sim 3 \%$ of our \texttt{o3} runs produce two hot Jupiters in a system. Figures~\ref{fig:2g_aeim} and \ref{fig:2g_disc} show one example. A massive core ($\sim 5\,\mathrm{M_{\oplus}}$) forms at an early stage at $\sim 0.1\,\mathrm{Myr}$, and undergoes runaway gas accretion and opens a gap (figure \ref{fig:2g_disc}, second panel). The rapid growth of a giant induces collisional accretion and formation of a second core that undergoes runway gas accretion followed by gap formation (figure \ref{fig:2g_disc}, third panel). These two planets then grow at the viscous supply rate (equation~\ref{eq:accdomain}) and end up as gas giants with masses $\sim 398\,\mathrm{M_{\oplus}}$ and $542\,\mathrm{M_{\oplus}}$. 

The two giants in this run did not evolve much in terms of their semimajor axes. However, stronger dynamical interactions between the two giants result in higher eccentricities and inclinations (figure \ref{fig:2g_aeim}, blue and red lines) than for single hot Jupiter systems (e.g figure~\ref{fig:dym_evo_aeim} and \ref{fig:WASP47o34k_aeim}, blue lines). 

Recent observations of the WASP-148 system show that it may have two close-in giants (both have mass $\sim 100\,\mathrm{M_{\oplus}}$) orbiting and transiting the same star \citep{2020A&A...640A..32H}. The semimajor axes of the two giants, WASP-148 b and c, are at $\sim 0.08$ and $0.21\,\mathrm{au}$ respectively. These two giants are on more eccentric orbits ($0.2<e<0.36$) than obtained in our simulation (see figure \ref{fig:2g_aeim}), and this may be because the two giants in our simulation did not experience strong scattering with another giant planet.

\section{Discussion and conclusions}\label{sec:discon}
In this study, we have presented the results of $N$-body simulations of the \textit{in situ} formation of planetary systems containing a hot Jupiter and super-Earths that orbit interior to the giant planet. The aim is to examine whether or not planetary systems with this architecture can form locally by the collisional accretion of planetary embryos followed by gas envelope accretion onto cores that grow to sufficient mass.

This study is motivated by observations that demonstrate the existence of planetary systems with the above described architecture. In particular, the systems WASP-47, Kepler-730, TOI-1130, Kepler-487, and Kepler-89 all have a transiting giant with an orbital period less than 30 days and contain at least one inner transiting super-Earth/mini-Neptune. We use these systems as templates to construct the initial conditions for the $N$-body simulations, where the orbital locations of the giant planets define the median semi-major axes of the planetary embryos. 

Two different sets of initial conditions were considered, a `seed-model' and an `equal-mass model'. In the seed-models, a relatively massive seed-protoplanet ($4$ or $4.5\,\mathrm{M_{\oplus}}$) was placed at the reference semi-major axis, with multiple $0.5\,\mathrm{M_{\oplus}}$ embryos orbiting interior and exterior to the seed. The equal-mass models have a chain of equal mass embryos ($0.5\,\mathrm{M_{\oplus}}$) centred at the reference semi-major axis. The purpose of the seed-model was to study the dynamical evolution of systems in which a giant planet was essentially guaranteed to form, while the equal-mass simulations investigate the giant planet occurrence rate using a set of unbiased initial conditions.
 
The $N$-body simulations included a realistic collision model and a protoplanetary disc modelled as an $\alpha$-disc subject to photoevaporation. The disc provided eccentricity/damping forces on the protoplanets that were also able to accrete gas from the disc. Orbital migration through disc-planet interactions was neglected.

We observed that the dynamical evolution of the seed-models consistently followed 4 phases of evolution: 1) an early impact phase, where embryo-embryo collisions occur frequently at early times because of the initial compact configurations of the embryos, and where the systems stabilize after this initial epoch of collisional evolution due to the disc damping forces; 2) a runaway gas accretion phase, where the seed-protoplanet undergoes runaway gas envelope accretion, and the rapid increase in the mass of this planet dynamically heats up other bodies in the system; 3) an outer disc damping phase, where the embryos exterior to the seed protoplanet continue to experience the disc damping forces and the interior planets experience almost no disc damping force because of gap formation; 4) a late chaotic phase, where giant impacts between planets are common due to the dispersal of the gas disc. 

By design, the seed-model is efficient at producing systems of coexisting hot Jupiters and inner super-Earths/sub-Neptunes. The final average multiplicity of the inner systems is 2.2, while the average outer system multiplicity is 3.5. The inner system planets are more massive than the outer planets on average. The average inner system planet mass is $13.2\,\mathrm{M_{\oplus}}$, and that of the outer system is $4.8\,\mathrm{M_{\oplus}}$. All seed-model runs result in a final system consisting of a hot Jupiter and inner super-Earths/mini-Neptunes (except for the \texttt{WASP47} templates for reasons discussed in section \ref{subsec:seedevo}).

The equal-mass models require efficient collisional accretion among the embryos to occur in order for a giant planet to form within the gas disc lifetime. Only $\sim1\%$ of the equal-mass runs produced a gas giant planet. The formation history for such systems follows the four phases described above for the seed-model. For the runs where no giant forms, the disc damping forces are dominant throughout the gas disc lifetime, and embryo collisions are only common at the very early stages of the simulations (due to the initial compact configurations) and after the gas disc disperses. 

For the equal-mass runs, we undertook synthetic transit observations of our final planetary systems, and the proportion of the detected systems that contained a hot Jupiter and at least one inner super-Earth was $0.26\%$, similar to the occurrence rate of $\sim0.2\%$ for such systems from actual transit surveys. However, we find that there is only a $\sim5\%$ chance that a hot Jupiter would be detected as a single planet without nearby interior or exterior super-Earths, and so it is clear that the model presented here cannot explain the majority of hot Jupiter systems. The final planetary systems we form containing a hot Jupiter and nearby super-Earths are too flat to agree with the results of transit surveys such as Kepler.

A crucial ingredient in the models is the gas envelope accretion prescription, and we examined the impact of adopting different gas accretion routines. This included the simple fit from \citet{2016MNRAS.457.2480C} to the 1D gas accretion simulations conducted at 5.2\,au by \citet{2010Icar..209..616M}, and a new fit to a large suite of 1D gas accretion simulations conducted at different orbital radii and with different envelope opacities using the 1D envelope structure model of \citet{2017MNRAS.470.3206C}. One conclusion from our study is that a significant opacity reduction factor of, $f_{\mathrm{opa}}$, is required to form a hot Jupiter in these simulations. A value of $f_{\mathrm{opa}}=10^{-2}$ is not small enough, whereas $f_{\mathrm{opa}}=10^{-3}$ results in approximately $6\%$ of the equal-mass simulations forming a hot Jupiter, which is higher than the $\sim1\%$ rate obtained when using the \citet{2016MNRAS.457.2480C} gas accretion prescription. Due to the high efficiency of giant formation with $f_{\mathrm{opa}}=10^{-3}$, $\sim3\%$ of the runs formed systems with two giant planets, similar to the recently reported planetary system, WASP-148 \citep{2020A&A...640A..32H}.

The simulations presented here show that within the parameter space that we have considered, the formation of systems containing a hot Jupiter and at least one interior super-Earth/mini-Neptune can form \textit{in situ} through collisional accretion in a compact chain of planetary embryos, followed by gas accretion onto a core that grows to be of sufficient mass to undergo runaway gas accretion. Overall, the frequency with which such systems are detected when we synthetically observe the simulation outcomes is similar to the frequency of occurrence in actual transit surveys. However, closer inspection of the distribution of system multiplicities arising from the simulations shows that they do not match the observations, as the models predict that hot Jupiters should be rarely be detected as single planets. 

The model presented here can form systems similar to WASP-47, Kepler-730, and TOI-1130, and may represent the means by which systems with these particular architectures formed. The failure to form hot Jupiter systems that appear as single planets in transit surveys, however, suggests an alternative formation scenario is required to explain the majority of hot Jupiter systems.

\section*{Acknowledgements}
The authors wish to thank Alesssandro Morbidelli and Seth Jacobson for making their version of the \textsc{symba} $N$-body code available, and an anonymous referee whose comments allowed us to significantly improve this paper. This research utilised Queen Mary's Apocrita HPC facility, supported by QMUL Research-IT\footnote{http://doi.org/10.5281/zenodo.438045}.
Richard Nelson acknowledges support from STFC through the Consolidated Grants ST/M001202/1 and ST/P000592/1. Richard Nelson and Gavin Coleman acknowledge support from the Leverhulme Trust through grant RPG-2018-418.
This research has made use of the NASA Exoplanet Archive, which is operated by the California Institute of Technology, under contract with the National Aeronautics and Space Administration under the Exoplanet Exploration Program.

\section*{Data availability}
The data underlying this article will be shared on reasonable request to the corresponding author.




\bibliographystyle{mnras}
\bibliography{Sanson} 



\appendix
\section{Mass-radius relation}
\begin{figure}
\begin{center}
\includegraphics[width=\columnwidth]{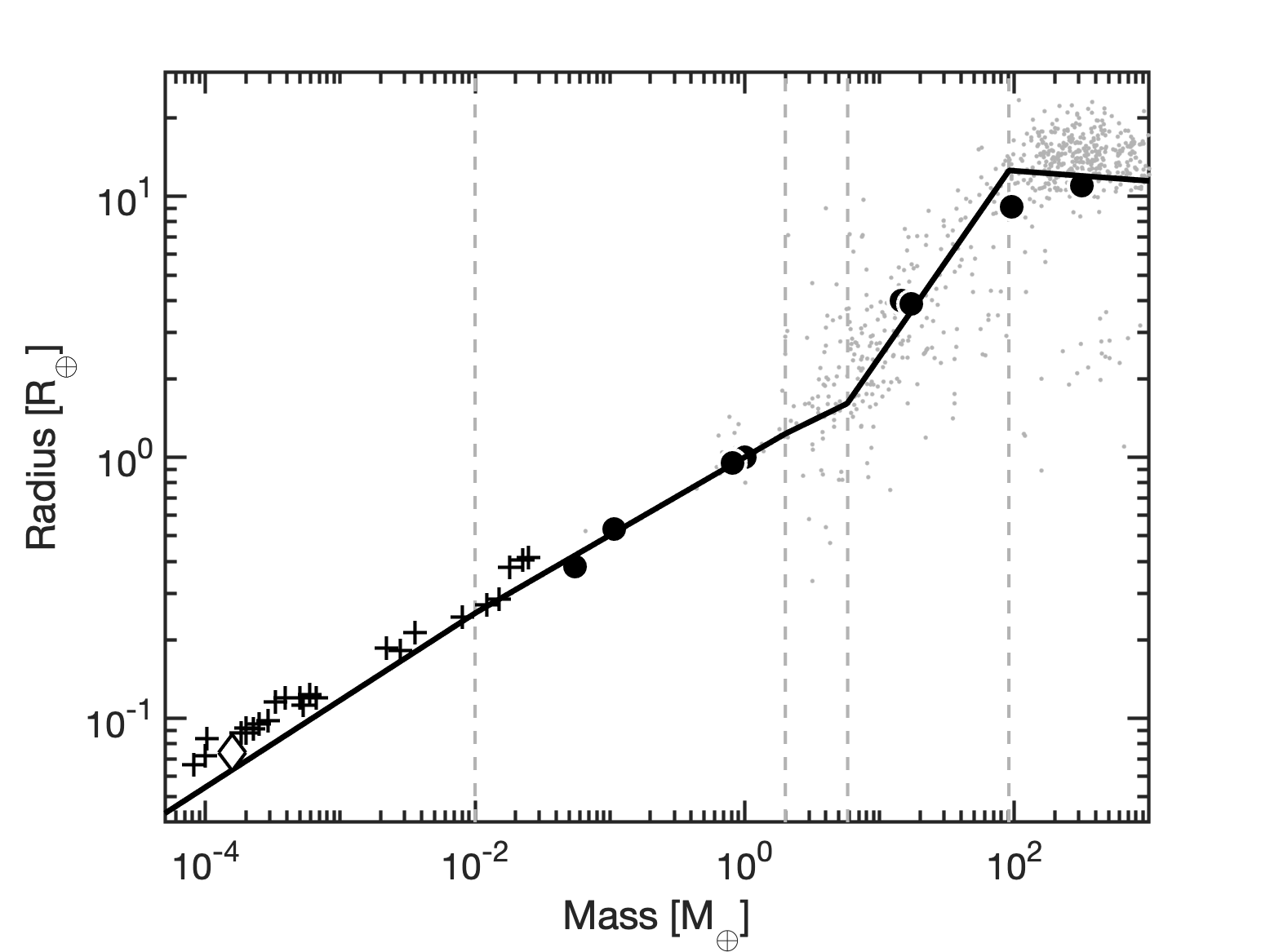}
 \caption[Mass-radius relation]{Mass-radius relation adopted in this study, as shown by equation \ref{eq:mass-radius}. Comparison to other astronomical objects is provided, including the Solar System planets (black dots), confirmed exoplanets (grey dots), other Solar System objects larger than $400\,\mathrm{km}$ ($+$), and Ceres ($\diamondsuit$). Grey vertical dashed lines denote the different planetary mass intervals.}
 \label{fig:mass-radius}
\end{center}
\end{figure}

The mass-radius relation that we adopted in this study is 
\begin{equation}\label{eq:mass-radius}
R_{\mathrm{p}}=c_1\left( \frac{M_{\mathrm{p}}}{\mathrm{M_{\oplus}}} \right)^{c_2}\mathrm{R_{\oplus}},
\end{equation}
where $c_1$ and $c_2$ are the mass-radius relation coefficients and are different in different domains. The values of $c_1$ and $c_2$ are listed in table \ref{tab:mass-radius}. Figure \ref{fig:mass-radius} shows the mass-radius relation, and also shows the locations of different astronomical objects in the mass-radius plane. These objects include exoplanets, Solar System planets, and satellites and minor bodies from the Solar System. We can see the adopted relation does a decent job of reproducing the various bodies shown. In the lowest mass interval the fit fails to match the dwarf planets plotted there (including Ceres) because they are icy whereas the fit assumes a lunar density. In the adjacent mass interval the fit passes through the Moon and the Earth, and hence reproduces these bodies by design.
\begin{table}
\caption[Mass-radius relation - $c_1$ and $c_2$]{The value of $c_1$ and $c_2$ applied to the mass-radius relation (equation \ref{eq:mass-radius}).
}\label{tab:mass-radius}
\begin{tabularx}{1.0\columnwidth}{C{1.5}R{1.0}R{0.5}}
\toprule\toprule
Planet mass range [$\mathrm{M_{\oplus}}$] & $c_1$ & $c_2$\\
\midrule  
$M_{\mathrm{p}}\leqslant0.01$ 		& 1.1748 & 0.3333\\
$0.01<M_{\mathrm{p}}\leqslant 2$ 	& 1.0014 & 0.2987\\
$2<M_{\mathrm{p}}\leqslant 5.8$ 	& 1.0340 & 0.2524\\
$5.8<M_{\mathrm{p}}\leqslant 91$	& 0.4372 & 0.7440\\
$91<M_{\mathrm{p}}\leqslant 1000$ 	& 14.9562 & -0.0386\\
\bottomrule\bottomrule
\end{tabularx}
\end{table}

\section{System parameters}
The system parameters of the five selected systems that we focused in this study, including WASP-47 \citep{2015ApJ...812L..18B}, Kepler-730 \citep{2019ApJ...870L..17C}, TOI-1130 \citep{2020ApJ...892L...7H}, Kepler-487 \citep{2016ApJ...822...86M}, and Kepler-89 \citep[also known as KOI-94,][]{2013ApJ...768...14W}.
\begin{table}
\caption[Stellar and planet parameters]{Stellar and planet parameters of the five selected systems. All planets in the table are transiting planets and the known masses are given in the form of actual mass, except WASP-47c which is detected by RV$^{\dagger}$ and the mass is in $M_{\mathrm{p}}\sin{I}$.$^{\ddagger}$ We list this planet here for the completeness, but the formation process of this cold giant is not the main consideration in this study. KOI-191.02 and 191.03 are Kepler candidates orbiting in the Kepler-487 (KOI-191) system.$^{\ast}$ }\label{tab:systemparameter}
\begin{tabularx}{1.0\columnwidth}{C{0.7}C{1.6}C{0.7}}
\toprule\toprule
& Stellar Parameters & \\
\cmidrule(r){2-2}
\end{tabularx}
\begin{tabularx}{1.0\columnwidth}{L{1.7}R{0.8}R{0.8}R{0.7}}
Host name & $M_{\star}~[\mathrm{M_{\odot}}]$ & $R_{\star}~[\mathrm{R_{\odot}}]$ & $T_{\star}~[\mathrm{K}]$ \\
\midrule
WASP-47 		& 1.040 & 1.137 & 5552 \\
Kepler-730 	& 1.047 & 1.411 & 5620 \\
TOI-1130 	& 0.684 & 0.687 & 4250 \\
Kepler-487 	& 0.910 & 0.880 & 5444 \\
Kepler-89 	& 1.277 & 1.520 & 6182 \\
\end{tabularx}
\begin{tabularx}{1.0\columnwidth}{C{0.7}C{1.6}C{0.7}}
\midrule
& Planet Parameters & \\
\cmidrule(r){2-2}
\end{tabularx}
\begin{tabularx}{1.0\columnwidth}{L{1.2}L{0.75}R{1.075}R{1.075}R{0.90}}
\multicolumn{2}{l}{Planet name} & $a$ [au] & $M_{\mathrm{p}}$ [M$_{\oplus}$] & $R_{\mathrm{p}}$ [R$_{\oplus}$]\\
\midrule
WASP-47 & e & 0.01694 & 6.83 & 1.81 \\
- 		& \textbf{b} & \textbf{0.05129} & \textbf{363.1} & \textbf{12.63}\\
- 		& d & 0.08600 & 13.1 & 3.576\\
- 		& c$^{\dagger}$ & 1.3926 & 398.2$^{\ddagger}$ & - \\ [5pt]
Kepler-730 	& c & 0.03997 & - & 1.57 \\
- 			& \textbf{b} & \textbf{0.0694} & - & \textbf{12.33}\\ [5pt]
TOI-1130 	& b & 0.04394 & - & 3.65 \\ 
- 			& \textbf{c} & \textbf{0.07098} & \textbf{309.6} & \textbf{16.8}\\  [5pt]
Kepler-487 	& 191.03$^{\ast}$ & 0.0149 & - & 1.20 \\
- 			& 191.02$^{\ast}$ & 0.0337 & - & 2.25 \\
- 			& \textbf{b} & \textbf{0.11719} & - & \textbf{11.42} \\
- 			& c & 0.21682 & - & 2.68\\ [5pt]
Kepler-89 	& b & 0.05119 & 10.5 & 1.71 \\
- 			& c & 0.1013 & 15.6 & 4.32\\
- 			& \textbf{d} & \textbf{0.1684} & \textbf{106.0} & \textbf{11.27}\\
- 			& e & 0.3046 & 35.0 & 6.56\\
\bottomrule\bottomrule
\end{tabularx}
\end{table}


\newpage 
\bsp	
\label{lastpage}
\end{document}